\documentclass{aa}
\usepackage{graphicx}

\usepackage{xcolor}
\usepackage{ulem}
\usepackage{comment}
\usepackage{natbib}
\usepackage{amssymb}
\usepackage{scrextend} 
\usepackage{placeins} 
\usepackage{hyperref}
\hypersetup{
    colorlinks=true,
    linkcolor=blue,
    filecolor=magenta,      
    urlcolor=blue,
    citecolor=blue,
    }
\usepackage{multirow}
\usepackage{txfonts}
\begin{document}

   \title{An observational study of rotation and binarity \\ of Galactic O-type runaway stars}

       \author{M. Carretero-Castrillo
          \inst{1,2,3}
          \and M. Rib\'o \inst{1,2,3}\fnmsep\thanks{Serra H\'unter Fellow}
          \and J. M. Paredes \inst{1,2,3}
          \and G. Holgado \inst{4,5}
          \and C. Martínez-Sebastián \inst{4,5}
          \and S. Simón-Díaz \inst{4,5}
          }

   \institute{Departament de Física Quàntica i Astrofísica (FQA), Universitat de Barcelona (UB),  c. Martí i Franquès, 1, 08028 Barcelona, Spain
              \and
              Institut de Ciències del Cosmos (ICCUB), Universitat de Barcelona (UB), c. Martí i Franquès, 1, 08028 Barcelona, Spain
              \and
              Institut d'Estudis Espacials de Catalunya (IEEC), Edifici RDIT, Campus UPC, 08860 Castelldefels (Barcelona), Spain
              \and
              Instituto de Astrofísica de Canarias, c/Vía Láctea, S/N, 38205 La Laguna, Tenerife, Spain
              \and
              Departamento de Astrofísica, Universidad de La Laguna, 38206 La Laguna, Tenerife, Spain\\
              \email{mcarretero@fqa.ub.edu}
             }

   \date{Received Received 29 July 2025; accepted 23 October 2025}

  \abstract
  {\textit{Gaia}~DR3 data have revealed new massive runaway stars, while large spectroscopic surveys now enable detailed characterization studies. However, the relative contributions of binary supernova (BSS) and dynamical ejection (DES) scenarios to explain their runaway origin remain poorly constrained, particularly in the Milky Way.}
  {We aim to characterize the largest sample of Galactic O-type runaway stars ever investigated through their kinematics, rotation, and binarity {with the ultimate objective of shedding light into their potential runaway origins.}}
   {We use the GOSC-\textit{Gaia}~DR3 catalog of normal and runaway stars, and IACOB spectroscopic information to build a sample with 214 O-type stars with available information about projected rotational velocities ($v \sin{i}$), and a subsample of 168 O-type stars with additional information about their likely single (LS) or single-lined (SB1) spectroscopic binary nature. We also consider an additional sample of 65 double-lined (SB2) spectroscopic binaries.}
   {We find that among our sample of Galactic O-type runaways, most (74\%) have $v \sin{i}<200$~km~s$^{-1}$, {whereas for normal stars this fraction is slightly higher (82\%)}. There are no fast-moving ($V_\text{PEC}^\text{2D}> 85$~km~s$^{-1}$) runaways being fast rotators ($v \sin{i}\geq200$~km~s$^{-1}$), except for \object{HD~124~979}. Runaways show lower SB1 fractions than normal stars, with no runaway SB1 fast-rotating systems; on average, runaways rotate faster than normal stars; and their runaway fraction is higher among fast rotators (44\%) vs. the slow rotators (34\%). This is consistent with BSS dominance for fast rotators. We also found that SB2 systems hardly reach runaway velocities with a low runaway fraction (10\%). Runaways with $V_\text{PEC}^\text{2D}>60$~km~s$^{-1}$ are mostly single and interpreted as DES products, while runaways with $V_\text{PEC}^\text{2D}>85$~km~s$^{-1}$ are also interpreted as two-step products, with the binary \object{V479~Sct}/\object{LS~5039} a likely example. Finally, three of 12 runaway SB1 systems are high-mass X-ray binaries.}
   {Our observational study reveals that Galactic O-type runaways are dominated by slow rotators. The study points to a dominance of BSS among fast-rotating runaways, and of DES and two-step among the high-velocity ones. The observed trends provide valuable constraints on models about runaway origins.}
   
   \keywords{binaries: spectroscopic -- 
             stars: early-type --
             stars: kinematics and dynamics --
             stars: rotation --
             supernovae: general --
             X-rays: binaries} 

   \maketitle
%

\section{Introduction} \label{Sec:Intro}

Runaway stars are characterized by their high peculiar velocity with respect to their surrounding environment \citep{Blaauw1961}.
Observationally, more than 20\% of the O-type stars and about 5--10\% of the B-type stars are found to be runaway stars \citep{Blaauw1961,Stone1979,Tetzlaff2011,MA2018,Kobulnicky2022,Guo2024}. In a recent work to search for runaway stars in the Milky Way, \cite{MCC2023} used \textit{Gaia}~DR3 data and identified runaway fractions of $\sim$25\% and $\sim$5\% for O and Be stars, respectively. Therefore, massive runaway stars constitute a significant fraction of massive stars.

Massive runaway stars impact both their host and neighboring clusters, providing insights into cluster dynamics and interactions (e.g., \citealt{Stoop2024}). They play an important role as sources of stellar feedback \citep{Larson1974}. Due to their high velocities, they travel significant distances from their birthplaces, interacting with the interstellar medium (ISM) and often producing stellar bow shocks \citep{NoriegaCrespo1997,Peri2015,Kobulnicky2016,Bodensteiner2018,MCC2025}. As a result, runaway stars escape their birth clouds, and thus their ionizing radiation is more likely to contribute to the ionization of the ISM \citep{Conroy2012,Kimm2014}. In addition, O-type stars are mostly found in binary systems, with at least 70\% of them expected to interact with a close-by companion at some point of their lives \citep{Chini2012,Sana2012,MoeStefano2017,Guo2022,Offner2023,MarchantBodensteiner2024}. These binaries can evolve into peculiar systems such high-mass X-ray binaries (HMXBs) \citep{Lewin2006}, millisecond pulsar systems, or double-degenerated binaries \citep{Heuvel2007}, which are potential sources of gravitational waves (e.g., \citealt{Langer2020}). In addition, the occurrence of runaway ejections in binary systems can alter their evolutionary pathways.

Two main scenarios have been proposed to explain the runaway origin of massive stars due to different ejection mechanisms: (1) the dynamical ejection scenario (DES, \citealt{Poveda1967,Portegies1999}), in which a massive star is ejected by a close three- or four-body interaction in the core of a dense cluster; and (2) the binary supernova scenario (BSS, \citealt{Blaauw1961,LeonardDuncan1988}), in which the kick by the supernova (SN) explosion in a binary system either ejects the surviving binary (now with a compact object) as a runaway, or disrupts the system, ejecting both components separately.
According to simulations of the BSS, most binaries are disrupted (e.g., \citealt{Eldridge2011,Renzo2019,Wagg2025}). However, there is observational evidence of runaway binaries with compact objects that can evolve into HMXBs and gamma-ray binaries \citep{BekensteinBowers1974,Stone1982,Oijen1989,Moldon2012,Marcote2018,MCC2023}. A two-step ejection process combining both scenarios, DES and BSS, might also occur \citep{Pflamm-Altenburg2010}. Although particular studies have been conducted in the Small Magellanic Cloud (SMC) and Large Magellanic Cloud (LMC) \citep{DorigoJones2020,Sana2022,Phillips2024}, the relative contributions of these scenarios in the Galaxy remain uncertain.

Massive O-type stars have been observed to exhibit a wide range of projected rotational velocities $(v \sin{i})$. Empirically, the $v \sin{i}$ distribution of O-type stars is characterized by a main component including stars with $v \sin{i}<100$~km~s$^{-1}$, and a tail of fast rotators with values reaching up to 400--600~km~s$^{-1}$. This has been found in different metallicity environments such us the Milky Way \citep{ContiEbbets1977,SimonDiaz2014,Holgado2022}, the LMC  \citep{RamirezAgudelo2013,RamirezAgudelo2015}, and the SMC \citep{Bodensteiner2023}, with \citet{PennyGies2009} providing a comparative study across these galaxies. The stellar spin can change due to close binary interaction through different mechanisms: by tides (e.g., \citealt{Zahn1975,Hut1981,deMink2009}), mass transfer (e.g., \citealt{Packet1981,Pols1991,Petrovic2005,Dervisoglu2010}), or when two stars merge (e.g., \citealt{Podsiadlowski1992,Tylenda2011}). Based on a fast population-synthesis model, \cite{deMink2013, deMink2014} suggested that binary interactions significantly influence the rotational velocity distribution of massive stars. In fact, as a result of their simulations, they found a large percentage of O-type stars with $v \sin{i}$ larger than $>200$~km~s$^{-1}$, a threshold commonly used to define fast-rotating stars, and interpreted them as post-interaction binary products. In addition, while mergers may contribute to the post-interaction binary population, they are likely slow rotators due to rapid magnetic braking in merger products \citep{Ferrario2009,Schneider2019}. All this suggests that the fast-rotating tail is populated predominantly by mass transfer products rather than mergers or stars born with such a high rotational velocity (see, however, \citealt{Naze2023}).

Runaway stars are expected to show distinct rotational signatures linked to their formation. On the one hand, the BSS may produce fast-rotating runaways through pre-SN mass transfer, since mass transfer spins up the surviving star and may enrich it with He/N-rich material \citep{Blaauw1993}. Growing observational evidence supports this connection between BSS and fast rotation. In this context, in a small sample limited to 700~pc, \cite{Hoogerwerf2001} found that massive runaways are fast-rotating and He-rich. In addition, \cite{MA2018} showed that Galactic O runaway stars rotate on average faster than the non-runaways. Moreover, \cite{Britavskiy2023} found a runaway fraction of $\sim$33--50\% among fast-rotating O-type stars vs. $\sim$20--30\% among slow-rotating ones. Finally, \cite{MartinezSebastian2025} found a higher percentage of single-lined spectroscopic binary (SB1) systems and runaways among He-enhanced O-type stars. On the other hand, runaways produced by DES, being either single or binaries before the ejection, do not need a previous mass-transfer period that would imply fast rotation. Therefore, the DES is not expected to be that efficient in producing runaway stars with fast rotation, which would only occur in rare cases where the ejected binaries also experience spin-up binary interactions (see introduction in \citealt{Hoogerwerf2001}). In addition, BSS and DES are expected to produce different peculiar velocity distributions based on simulations, with the DES runaway population showing larger velocities and higher masses (e.g., \citealt{Poveda1967,LeonardDuncan1988,PeretsSubr2012,Renzo2019}). For the 30 Dor region of the LMC, \cite{Sana2022} found two populations of runaway stars, a fast-rotating but slow-moving one, and a slow-rotating but fast-moving one, that were interpreted as produced by BSS and DES, respectively. As for the binary fraction, as already mentioned, most BSS runaways are expected to become unbound according to simulations, whereas DES produces singles and binaries, with the higher peculiar velocity ones typically being singles \citep{LeonardDuncan1988,Hoogerwerf2001,Eldridge2011,PeretsSubr2012,Oh_Kroupa_2016,Renzo2019}. In particular, double-lined spectroscopic binary (SB2) systems act as direct tracers of DES. Therefore, the study of the binary fraction of massive runaway stars is also important to understand the pre- and post-runaway stages.

A comprehensive understanding of the various possible dynamical origins of runaway stars requires the combined analysis of kinematics (space velocities and rotational velocities) and binarity for both normal and runaway massive star populations. This approach can shed light on the runaway ejection mechanisms but can also help to constrain the initial conditions of the parent clusters and of massive binary evolution. Building on our catalog of O runaway stars based on \textit{Gaia}~DR3 \citep{MCC2023}, in this work we incorporate spectroscopic information from the IACOB project \citep{SimonDiaz2014,Holgado2022}.  To the best of our knowledge, this work represents the most extensive observational study to date of kinematics, rotation, and binarity of Galactic O-type stars with accurate runaway classifications.

This paper is organized as follows. In Sect.~\ref{Sec:Catalogs} we describe the different input data we used and the two samples we compiled for the analysis. In Sect.~\ref{sec:results} we present the results we obtained for the kinematics and the binarity study. In Sect.~\ref{sec:discussion} we discuss the velocity distributions obtained for normal and runaway stars, interpret them in terms of the runaway ejection scenarios, and comment on possible binary candidates to HMXBs. We summarize our main findings and provide an outlook in Sect.~\ref{Sec:Conclusions}.

\section{Data} \label{Sec:Catalogs}

\subsection{GOSC-\textit{Gaia} DR3 catalog} \label{sec:GOSCGaiaDR3}

In \cite{MCC2023}, we searched for Galactic O runaway stars using \textit{Gaia}~DR3 astrometric data\footnote{We also searched for Be runaway stars. Both the O and Be runaway catalogs obtained in \cite{MCC2023} are available at the \href{https://cdsarc.cds.unistra.fr/viz-bin/cat/J/A+A/679/A109}{CDS}.}. For this, we first cross-matched the Galactic O-Star Catalog (GOSC; \citealt{GOSC}) with \textit{Gaia}~DR3 data. We also applied several quality cuts, and produced the GOSC-\textit{Gaia}~DR3 catalog with 417 O-type stars. This catalog contains 106 O runaway stars (42 new identifications), corresponding to a runaway fraction of 25.4\%. The runaway stars were identified through an $E$ parameter (see Eq.~3 in \citealt{MCC2023}), which indicates the normalized significance at a 3-sigma confidence level of the runaway nature based on their 2-Dimensional (2D) velocities $(V_\text{PEC}^\text{2D})$ with respect to the mean Galactic rotation. In this way, stars with $E\ge1$ were classified as runaways, while stars with $E<1$ were classified as \textit{normal} stars\footnote{Note that in \cite{MCC2023} the authors called \textit{field stars} to those with $E<1$, but here we change the nomenclature to \textit{normal stars}, since this term is more commonly used.}. Therefore, higher $E$ values denote more confident runaway star identifications. This self-consistent methodology avoids using runaway velocity thresholds in contrast to the vast majority of previous works, and allows for the coherent identification of runaways, including stars moving at 2D peculiar velocities as small as $V_\text{PEC}^\text{2D}\sim16$~km s$^{-1}$. The runaways with 3D peculiar velocities $V_\text{PEC}^\text{3D}<30$~km s$^{-1}$ have been referred to as walkaways in the literature \citep{deMink2012, Renzo2019}. In particular, they have been defined in a theoretical context as those runaways coming from disrupted binaries after the SN explosion of the primary and former companion. In this work, we use an observational definition for the walkaways in our sample, as those identified as runaways in \cite{MCC2023} ($E>1$), but that have $V_\text{PEC}^\text{2D}<25$~km~s$^{-1}$. This is a commonly used 2D runaway velocity threshold (e.g., \citealt{Kobulnicky2022}), which corresponds to the scaled 3D value of 30~km~s$^{-1}$ to 2D ($30\sqrt{2/3}\simeq25$).

\subsection{Spectroscopic information from the IACOB project} \label{sec:IACOBdata}

We took advantage of the long-term efforts made by the IACOB project (P.I. Sim\'on-D\'iaz) to compile an extensive catalogue of empirical properties of Galactic massive OB-type stars, from high-quality spectroscopic observations\footnote{\href{https://research.iac.es/proyecto/iacob/iacobcat/}{https://research.iac.es/proyecto/iacob/iacobcat/}} (see, e.g., \citealt{SimonDiaz2014,SimonDiaz2017,Holgado2018,Holgado2020,Holgado2022,deBurgos2024a,deBurgos2024b,deBurgos2025, MartinezSebastian2025}). Specifically, we benefited from the information on projected rotational velocities and detected signatures of spectroscopic binarity available -- either  publicly or through its internal catalogue -- for approximately 530 Galactic O-type stars.

As described in \cite{Holgado2022}, only stars not identified as SB2 have projected rotational velocity estimations (360 O-type stars in total). While some of the stars considered in this study were not yet included in the sample analyzed by \cite{Holgado2022}, we applied a similar analysis strategy. 
In particular, the measurements were obtained using the {\sc iacob-broad} tool \citep{SimonDiaz2014}, which efficiently disentangles the contributions of rotational and macroturbulent broadening to the observed line profiles. We consider uncertainties associated with the $v \sin{i}$ estimates on the order of 10\,\%.

Regarding the detection of signatures of binarity, we benefited from the multi-epoch nature of the IACOB spectroscopic database \citep[last described in][]{SimonDiaz2020}, which includes a minimum of three spectra for more than 75\% of the sample of O-type stars. 
The stars were classified as double-line spectroscopic binaries (SB2) if the line of He~{\sc i}$\,\lambda$5875 shows two components in any of the available spectra.
Otherwise, they were classified as single-line spectroscopic binaries (SB1) or likely-single (LS) based on radial velocity (RV) variability following a combination of criteria presented in \cite{SimonDiaz2024} and \cite{Britavskiy2023} -- for the case of stars with $v \sin{i}$ below and above $\sim$200\,km\,s$^{-1}$, respectively. This classification strategy accounts for the intrinsic variability of O-type stars, which varies in amplitude depending on their stellar parameters.
We applied a Markov Chain Monte Carlo approach to the measured RVs, surface gravities, and effective temperatures to quantify the likelihood of misclassifying intrinsically variable LS stars as SB1 systems.
Typically, we required at least three spectroscopic observations to reliably determine binarity status, but we identified particular cases with fewer epochs when evidence of binarity was clear. Also, we note that some systems currently classified as LS may eventually be reclassified as SB1 as additional observational epochs become available.

\subsection{Cross-match and samples used in this work} \label{sec:samples}

A cross-match of the O-type stars included in both the GOSC-\textit{Gaia} DR3 and the IACOB resulted in 296 stars, which are both normal and runaway stars. From that initial sample, we removed 17 peculiar stars with spectral classifications such as Oe, Of?p, and extreme supergiants (\citealt{Negueruela2004,MA2024}), as well as another sample of 65 stars clearly identified as SB2\footnote{Although excluded of this sample, we will also comment on these SB2 systems in Sect.~\ref{sec:discussion_scenarios}}. We hence ended up with a \textit{clean} sample of non-peculiar LS and SB1 stars comprising 214 targets. For all the stars in this cleaned sample we have information available about their projected rotational velocities. Therefore, we call this sample the "rotation sample". In Table~\ref{Tab:rotsample} we present information of the first 10 stars of the rotation sample with the higher values of $E$ in decreasing order, while the full version is available at the CDS. Of the stars in the rotation sample we have information on their LS or SB1 nature for 168 of them (due multiepoch spectroscopy, see Sect.~\ref{sec:IACOBdata}), constituting a subsample that we call the "rotation LS-SB1 subsample".

\section{Results}\label{sec:results}

\subsection{Rotation sample}

As indicated in Sect.~\ref{sec:samples}, this sample contains 214 O-type stars with available $v \sin{i}$ measurements, of which 136 are normal and 78 are runaway stars. The runaway stars represent a percentage of 36.4\%. In this section we study the space and projected rotational velocities of the rotation sample.

\subsubsection{Projected rotational vs. space velocities} \label{sec:results_2DrotS}

\begin{figure*}
    \centering
    \includegraphics[width=0.97\hsize]{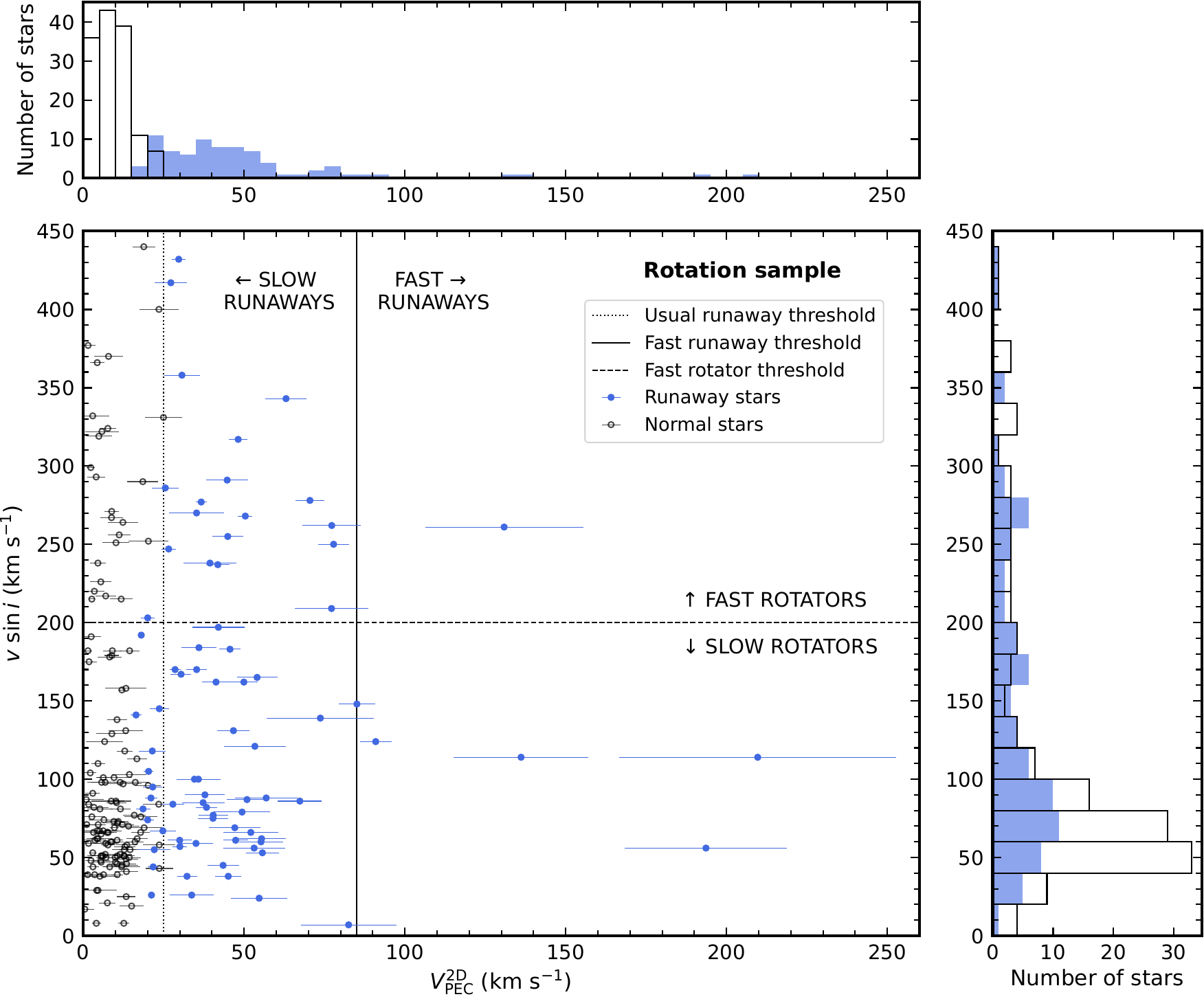}
    \caption{Projected rotational velocity as a function of the 2D peculiar velocity for the 214 O-type normal (black) and runaway (blue) stars in the rotation sample. The error bars in $V_\text{PEC}^\text{2D}$ are the individual uncertainties of the stars computed in \cite{MCC2023}. The uncertainties in $v \sin{i}$, not shown for clarity, are 10\% of the values. The vertical dotted and solid lines indicate the usual 2D runaway threshold and the fast runaway threshold, at 25 and 85~km~s$^{-1}$, respectively. The horizontal dashed line indicates the fast rotation threshold at 200~km~s$^{-1}$. \textit{Top}: $V_\text{PEC}^\text{2D}$ distribution for the normal and runaway stars with a bin size of 5~km~s$^{-1}$. \textit{Right}: $v \sin{i}$ distribution for the normal and runaway stars with a bin size of 20~km~s$^{-1}$.}
    \label{Fig:2D_rotsample}
\end{figure*}

Figure \ref{Fig:2D_rotsample} presents the projected rotational velocities as a function of the 2D peculiar velocities for the 214 stars in the rotation sample.  
There are three lines in the plot that we use to define different regions. First, the vertical dotted line indicates the commonly used 2D runaway threshold of 25~km~s$^{-1}$ (e.g., \citealt{Kobulnicky2022}). We show this threshold for illustrative purposes, since the runaway stars found in \cite{MCC2023} were identified without using a specific runaway threshold. Therefore, there are runaway stars in Fig.~\ref{Fig:2D_rotsample} to the left of this dotted line, which are those classified here as walkaway stars (see Sect.~\ref{sec:GOSCGaiaDR3}). Second, the vertical solid line indicates the 2D fast runaway threshold, located at 85~km~s$^{-1}$, which marks the transition between slow-moving and fast-moving runaway stars in 2D peculiar velocity. {To compute this threshold, we first scaled the 1D fast runaway threshold of 60~km~s$^{-1}$ from \cite{Sana2022} to 2D $\left(\text{by}~\hspace{-1mm}\sqrt{2/1}\right)$, and obtained $\sim$85~km~s$^{-1}$. We also obtained the weighted average of $V_\text{PEC}^\text{2D}$ for all the original 106 O-type runaway stars of \cite{MCC2023}, with weights computed as the inverse of the uncertainties in $V_\text{PEC}^\text{2D}$. Then, we obtained the fast runaway threshold as the value that deviates $2\sigma$ and $3\sigma$ from this weighted average, and decided to keep the $2\sigma$ one since it agrees with the scaled 2D fast runaway threshold of \cite{Sana2022}\footnote{We note that using the 78 runaway stars in the rotation sample, the 2D fast runaway threshold would be placed at $\sim$90~km~s$^{-1}$.}}. Finally, the horizontal dashed line indicates the fast-rotating threshold at 200~km~s$^{-1}$, inferred from mass transfer binary interaction simulations \citep{deMink2013}. Thus, in a similar way as done by \cite{Sana2022}, the fast runaway and rotating thresholds define four regions in Fig.~\ref{Fig:2D_rotsample}.

Most O-type normal and runaway stars are located in the slow-moving and slow-rotating region (see bottom-left region of  Fig.~~\ref{Fig:2D_rotsample}). The slow-moving and fast-rotating region is also populated by both normal and runaway stars. The fast-moving and slow-rotating region only contains 5 runaway stars (with one at the limit). Finally, the fast-moving and fast-rotating region contains only one runaway star: \object{HD~124~979}. In addition, all but one walkaway stars are slow rotators, with the exception at the limit.

Table~\ref{Tab:rotSnormRun} presents the number and percentage of normal and runaway stars and their subdivision in slow and fast rotators. We find 64\% of normal stars versus 36\% of runaway stars. Among normal stars we find 82\% of slow rotators and 18\% of fast rotators. These percentages are 74\% and 26\% for runaway stars, respectively. Therefore, although the percentage of fast rotators increases when considering runaway stars, the majority of Galactic O-type runaway stars are slow rotators. {This is expected given that most normal O-type stars are slow rotators.} Table~\ref{Tab:rotSslowFast}  presents a complementary vision dividing first in slow and fast rotators and then subdividing in normal and runaway stars. We note here that the runaway fraction increases significantly when moving from slow to fast rotators (from 34 to 44\%), reaching almost half the sample. We emphasize that these numbers and percentages correspond to the largest observational data set ever studied of Galactic O-type runaway stars in space velocities and rotational velocities. Therefore, from the study of the rotation sample of Galactic O-type stars we can conclude that:

\begin{enumerate}[i)]
    \item Most runaways are slow rotators.
    \item Virtually all walkaways are also slow rotators.
    \item Fast-moving, fast-rotating runaways are practically absent.
\end{enumerate}

\renewcommand{\arraystretch}{1.1}
\begin{table}[]
\caption{Number and percentage for normal and runaway stars in the rotation sample, subdivided in slow and fast rotators.}
\label{Tab:rotSnormRun} 
\centering  
\resizebox{0.5\textwidth}{!}{
\begin{tabular}{cccc}
\hline \hline \vspace{-3mm}\\
\multicolumn{2}{c}{Normal Stars} & \multicolumn{2}{c}{Runaway Stars} \\
\hline \vspace{-3mm}\\
\multicolumn{2}{c}{136 (64\%)}   & \multicolumn{2}{c}{78 (36\%)} \\
\hline \vspace{-3mm}\\
Slow Rotators & Fast Rotators    & Slow Rotators & Fast Rotators \\
\hline \vspace{-3mm}\\
111 (82\%)    & 25 (18\%)        & 58 (74\%)     & 20 (26\%) \\
\hline
\end{tabular}
}
\end{table}

\renewcommand{\arraystretch}{1.1}
\begin{table}[]
\caption{Number and percentage for slow and fast stars in the rotation sample, subdivided in normal and runaway stars.}
\label{Tab:rotSslowFast} 
\centering  
\resizebox{0.5\textwidth}{!}{
\begin{tabular}{cccc}
\hline \hline \vspace{-3mm}\\
\multicolumn{2}{c}{Slow Rotators} & \multicolumn{2}{c}{Fast Rotators} \\
\hline \vspace{-3mm}\\
\multicolumn{2}{c}{169 (79\%)}    & \multicolumn{2}{c}{45 (21\%)} \\
\hline \vspace{-3mm}\\
Normal Stars & Runaway Stars      & Normal Stars & Runaway Stars \\
\hline \vspace{-3mm}\\
111 (66\%)   & 58 (34\%)          & 25 (56\%)    & 20 (44\%) \\
\hline
\end{tabular}
}
\end{table}

\subsubsection{Space velocities} \label{sec:results_vpec2d}

The $V_\text{PEC}^\text{2D}$ distribution for the 214 O normal and runaway stars in the rotation sample is presented in the top panel of Fig.~\ref{Fig:2D_rotsample}. We used a bin size of 5~km~s$^{-1}$ considering that the median of the individual uncertainties in $V_\text{PEC}^\text{2D}$ is $\sim$3.5~km~s$^{-1}$. Normal stars present a nearly flat distribution with a pronounced drop above 15~km~s$^{-1}$ and a few objects up to 25~km~s$^{-1}$. In contrast, runaway stars show a broad distribution above 15~km~s$^{-1}$ (the two possible broad maxima at $\sim$25 and $\sim$45~km~s$^{-1}$ are not statistically significant due to the space velocity uncertainties). In addition, in Fig.~\ref{Fig:vpec2d_run_norm} of Appendix~\ref{sec:App_SpaceVelocities} we show the distribution for the rotation sample together with the distributions for the fast-rotating and the slow-rotating sets, subdividing them into normal and runaway stars. 

For the normal stars the median of the $V_\text{PEC}^\text{2D}$ distributions when we consider the entire rotation sample, the fast, and the slow rotators, are $\sim$9, $\sim$8 and $\sim$9~km~s$^{-1}$, respectively. The $V_\text{PEC}^\text{2D}$ 95th-percentiles are $\sim$19, $\sim$23, and $\sim$18~km~s$^{-1}$, respectively. For the runaway stars, the median of the $V_\text{PEC}^\text{2D}$ for the same subsets as above are $\sim$41, $\sim$43, and $\sim$40~km~s$^{-1}$, respectively. The 95th-percentiles are $\sim$97, $\sim$80, and 98~km~s$^{-1}$, respectively. Therefore, the $V_\text{PEC}^\text{2D}$ distribution for the slow-rotating runaway stars is more spread out, but the fast-rotating runaway stars have a slightly higher median.

We show in Fig.~\ref{Fig:vpec2d_slow_fast}-left the $V_\text{PEC}^\text{2D}$ distributions for the entire rotation sample, the runaway and normal stars, subdividing them into fast- and slow-rotating sets. We show in Fig.~\ref{Fig:vpec2d_slow_fast}-right the corresponding cumulative distribution functions (CDFs), since they help to visualize possible differences between the distributions. We also show there in light colors the corresponding empirical cumulative distribution functions \citep{Glivenko1933}. We used the Kolmogorov-Smirnov (KS) test \citep{Smirnov1939} to compare the CDFs and found no statistically significant differences between the distributions (see results and further discussion in Appendix~\ref{sec:App_SpaceVelocities}).

\subsubsection{Projected rotational velocities} \label{sec:results_vsini_rotS}

The distribution of projected rotational velocities for the 214 O runaway and normal stars in the rotation sample is presented in the right panel of Fig.~\ref{Fig:2D_rotsample}. We used a bin size of 20~km~s$^{-1}$ considering that the median of the individual uncertainties in $v \sin{i}$ is $\sim$8~km~s$^{-1}$. For the normal stars, the distribution reaches a maximum around $\sim$40--80~km~s$^{-1}$, and shows a flat tail of fast rotators from $\sim$120~km~s$^{-1}$ up to $\sim$450~km~s$^{-1}$. For the runaway stars, it appears to have also a main maximum but now shifted toward $\sim$60--100~km~s$^{-1}$. This maximum is followed by a relatively flat tail but, in contrast with the normal stars, with possible small maxima around $\sim$170 and $\sim$270~km~s$^{-1}$. We tested bin sizes of 30 and 50~km~s$^{-1}$ and obtained similar results except for the disappearance of the maximum around $\sim$170~km~s$^{-1}$. However, the maxima around $\sim$60--100~km~s$^{-1}$ and $\sim$270~km~s$^{-1}$ are always present.

\begin{figure}
    \centering
    \includegraphics[width=\hsize]{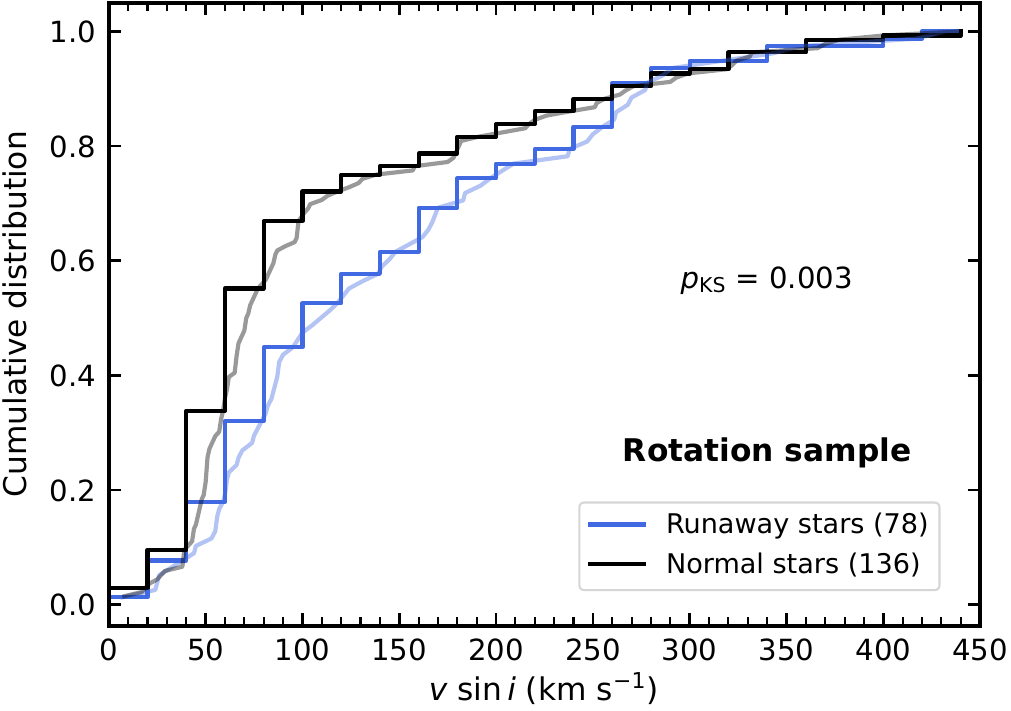}
    \caption{Cumulative distribution functions of $v \sin{i}$ for the 136 O normal (black) and 78 runaway stars (blue) in the rotation sample with bins of $\sim$20~km~s$^{-1}$. 
    The curves in light colors show the corresponding empirical cumulative distribution functions. $p_{\rm KS}$ is the corresponding $p$-value resulting from the KS test.}
    \label{Fig:CDF_vsini_rotS}
\end{figure}

For the normal stars, the $v \sin{i}$ distribution has a median (mean) of $\sim$72 ($\sim$115)~km~s$^{-1}$, while the 95th-percentile is $\sim$326~km~s$^{-1}$. For the runaway stars, the median (mean) is $\sim$114 ($\sim$144)~km~s$^{-1}$, and the 95th-percentile is $\sim$321~km~s$^{-1}$. The median and means already indicate that the $v \sin{i}$ distributions for normal and runaway stars are different. Fig.~\ref{Fig:CDF_vsini_rotS} shows the CDFs of the $v \sin{i}$ for normal and runaway stars in the rotation sample. In this case, the KS test indicates that the probability that the distributions come from the same parent distribution is 0.3\%. Therefore, they are statistically different. These differences come mainly from the range $\sim$50--250~km~s$^{-1}$.

\subsection{Rotation LS-SB1 subsample}\label{sec:rotbinsample}

As explained in Sect.~\ref{sec:samples}, the rotation LS-SB1 subsample contains 168 O-type stars with available $v \sin{i}$ measurements and LS or SB1 classification, of which 103 are normal and 65 are runaway stars. The runaway stars represent a percentage of 38.7\%, which is similar to the 36.4\% found in the rotation sample. In this section we study the binarity together with the projected rotational and 2D peculiar velocities for the rotation LS-SB1 subsample, and also comment on the runaway SB1 systems we identified.

\begin{figure*}
    \centering
    \includegraphics[width=0.8\hsize]{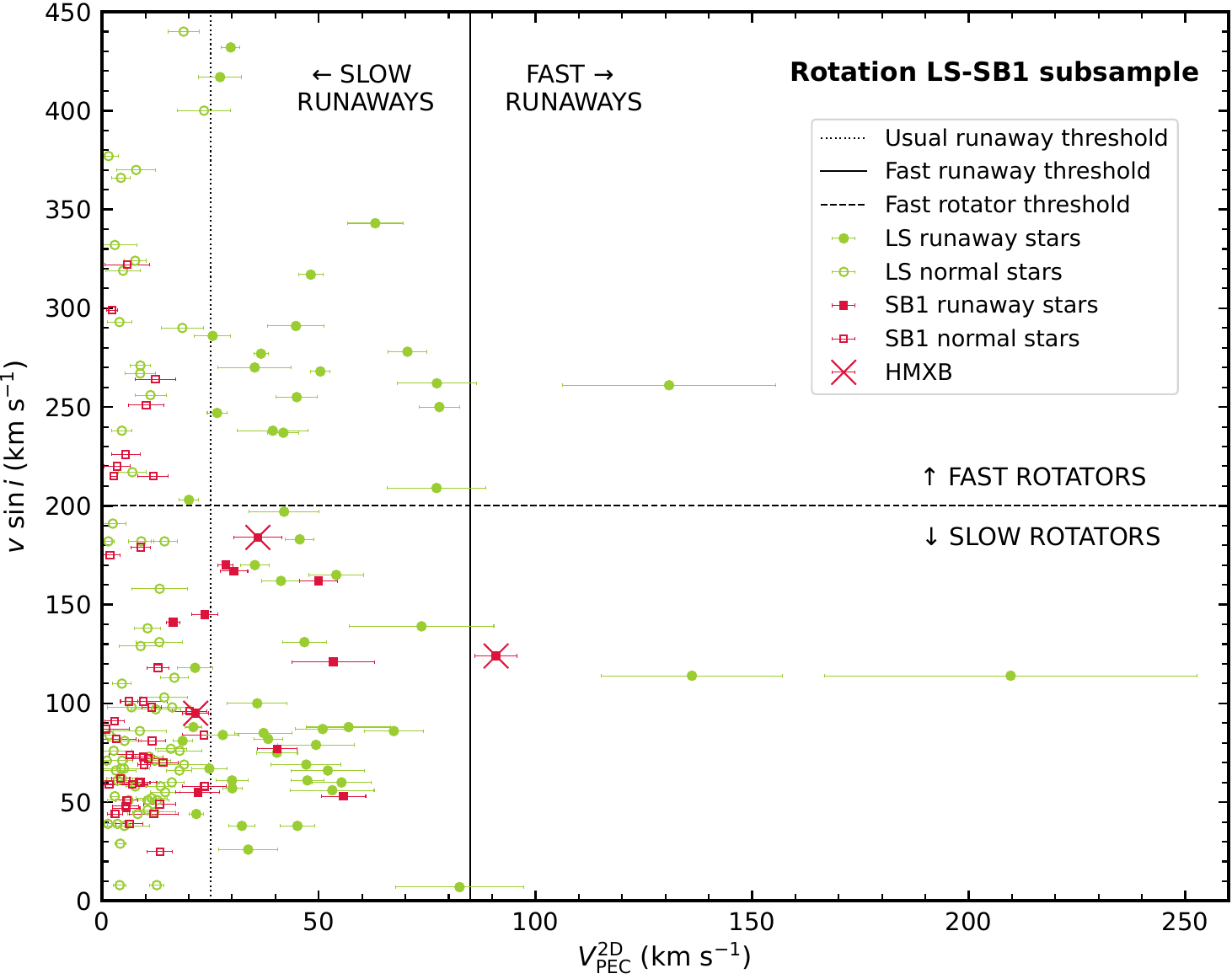}
    \caption{Projected rotational velocity as a function of the 2D peculiar velocity for the 168 O-type normal and runaway stars in the rotation LS-SB1 subsample. LS systems are shown as green circles, while SB1 systems are shown as red squares. Normal and runaway stars are represented with empty and filled symbols, respectively. The three known HMXBs identified among SB1 systems are indicated with red crosses. The uncertainties of the data points, and vertical and horizontal lines are the same as in Fig.~\ref{Fig:2D_rotsample}.}
    \label{Fig:2D_rotLS_SB1}
\end{figure*}

\subsubsection{Projected rotational vs. space velocities} \label{sec:results_rotLS_SB1}

In Fig.~\ref{Fig:2D_rotLS_SB1} we present a similar version of Fig.~\ref{Fig:2D_rotsample}, but now including the binarity status for the 168 O normal and runaway stars in the rotation LS-SB1 subsample. We also indicate the known 
HMXBs from the \cite{Fortin2023} catalog. We have lost about $\sim$20\% of the stars from the rotation sample, but still the rotation LS-SB1 subsample represents the largest observational sample studied in $v \sin{i}$, $V_\text{PEC}^\text{2D}$, and binarity for Galactic O-type runaway stars. The study of this subsample actually leads us to some exciting findings. There are no fast-rotating runaway SB1 systems, i.e, all fast-rotating runaways are likely single. This is in contrast with slow-rotating runaways, where we find both LS and SB1 systems, with percentages of 74\% and 26\%, respectively, as can be seen in Table~\ref{Tab:percent_rotLSSB1_regs}. It is also in contrast with the fast-rotating normal stars, where again we find both LS and SB1 systems, with percentages of 65\% and 35\%, respectively (see Table~\ref{Tab:percent_rotLSSB1_regs}). This discards any possible issue related to the SB1 identification in the fast-rotating domain. If we now focus on the slow-rotating SB1 systems, we find that their percentage decreases from 39\% to 26\% when moving from normal to runaway stars; in the case of fast rotators, as mentioned above, their percentage decreases from 35\% to 0\%. Therefore, another interesting result is that the fraction of SB1 systems is always lower (or even zero) for the runaway stars compared to the normal stars. We comment on the interpretation of these relevant results in Sect.~\ref{sec:discussion}.

\renewcommand{\arraystretch}{1.1}
\begin{table*}[t]
\caption{Number and percentage for normal and runaway stars in the rotation LS-SB1 subsample, subdivided in slow and fast rotators, and further subdivided in LS and SB1.}
\label{Tab:percent_rotLSSB1_regs}
\centering 
\begin{tabular}{cccc@{~~~~~~~}cccc}
\hline \hline \vspace{-3mm}\\
\multicolumn{4}{c}{Normal Stars} & \multicolumn{4}{c}{Runaway stars} \\
\hline \vspace{-3mm}\\
\multicolumn{4}{c}{103 (61\%)}   & \multicolumn{4}{c}{65 (39\%)} \\
\hline \vspace{-3mm}\\
\multicolumn{2}{c}{Slow rotators}  & \multicolumn{2}{c}{Fast rotators} & 
\multicolumn{2}{c}{Slow rotators}  & \multicolumn{2}{c}{Fast rotators} \\ 
\hline \vspace{-3mm}\\
\multicolumn{2}{c}{80 (78\%)} & \multicolumn{2}{c}{23 (22\%)} &
\multicolumn{2}{c}{46 (71\%)} & \multicolumn{2}{c}{19 (29\%)} \\
\hline \vspace{-3mm}\\
LS & SB1 & LS & SB1 & LS & SB1 & LS & SB1 \\
\hline \vspace{-3mm}\\
49 (61\%) & 31 (39\%) & 15 (65\%) & 8 (35\%) & 34 (74\%) & 12 (26\%) & 19 (100\%) & 0 (0\%) \\
\hline
\end{tabular}
\end{table*}

\renewcommand{\arraystretch}{1.1}
\begin{table*}[t]
\caption{Number and percentage for normal and runaway stars in the rotation LS-SB1 subsample, subdivided in LS and SB1, and further subdivided in slow and fast rotators.}
\label{Tab:percent_rotLSSB1}
\centering 
\resizebox{\textwidth}{!}{
\begin{tabular}{cccc@{~~~~~~~}cccc}
\hline \hline \vspace{-3mm}\\
\multicolumn{4}{c}{Normal Stars} & \multicolumn{4}{c}{Runaway stars} \\
\hline \vspace{-3mm}\\
\multicolumn{4}{c}{103 (61\%)}   & \multicolumn{4}{c}{65 (39\%)} \\
\hline \vspace{-3mm}\\
\multicolumn{2}{c}{LS}   & \multicolumn{2}{c}{SB1} & 
\multicolumn{2}{c}{LS}  & \multicolumn{2}{c}{SB1} \\ 
\hline \vspace{-3mm}\\
\multicolumn{2}{c}{64 (62\%)} & \multicolumn{2}{c}{39 (38\%)} &
\multicolumn{2}{c}{53 (82\%)} & \multicolumn{2}{c}{12 (18\%)} \\
\hline \vspace{-3mm}\\
Slow Rotators & Fast Rotators & Slow Rotators & Fast Rotators &
Slow Rotators & Fast Rotators & Slow Rotators & Fast Rotators \\
\hline \vspace{-3mm}\\
49 (77\%) & 15 (23\%) & 31 (79\%) & 8 (21\%) & 34 (64\%) & 19 (36\%) & 12 (100\%) & 0 (0\%) \\
\hline
\end{tabular}
}
\end{table*}

In a complementary way, in Table~\ref{Tab:percent_rotLSSB1} we present numbers and percentages of objects in the rotation LS-SB1 subsample for normal and runaway stars, subdivided in LS and SB1, and further subdivided in slow and fast rotators. The percentage of SB1 systems decreases from 38\% to 18\% when moving from normal to runaway stars. For normal LS stars we have 77\% of slow rotators and 23\% of fast rotators. The percentages for normal SB1 stars are quite similar, with 79\% of slow rotators and 21\% of fast rotators. The situation changes when we move to runaway stars: for LS the percentage of slow rotators slightly decreases to 64\% while the fast rotators represent 36\%; in contrast, for SB1 all of them are slow rotators. As mentioned above, the absence of runaway SB1 systems within fast rotators is discussed in Sect.~\ref{sec:discussion}.

Despite the poor statistics for the walkaway stars in the rotation LS-SB1 subsample (only ten stars) it is also worth mentioning that there are both LS and SB1 systems, with similar fractions.

Therefore, from the study of the rotation LS-SB1 subsample of Galactic O-type stars we can conclude that:

\begin{enumerate}[{i)}]
    \setcounter{enumi}{3}
    \item Runaways have reduced SB1 fractions versus normal stars.
    \item There are no runaway SB1 systems within fast rotators.
    \item In walkaways we find both LS and SB1 systems.
\end{enumerate}

In Appendix~\ref{sec:App_CDF_rotbinS} we present the CDFs in $v \sin{i}$ for the normal and runaway stars including the information on binarity (see Fig.~\ref{Fig:CDFvsini_LS_SB1}). We define four different sets: LS normal and runaway stars and SB1 normal and runaway systems. We also show the results of the KS tests between all these sets. While some comparisons could indicate a different origin of those distributions, these differences are not statistically significant. A summary of the results is included in Table~\ref{Tab:KStest_LS_SB1}.

\subsubsection{Runaway SB1 systems} \label{sec:SB1runaways}

In this section we comment on the runaway SB1 systems that we found and we present an update on their multiwavelength properties. The unseen companions in SB1 systems could be main-sequence stars, stripped helium stars, or compact objects such as neutron stars (NSs) or stellar-mass black holes (BHs). In the context of runaways, the case with compact objects would correspond to BSS products. If these systems with a massive star and a compact object are close enough, they can evolve through a second mass-transfer phase during which they can be detected as HMXBs. Indeed, three of our runaways are HMXBs: \object{V479~Sct}, \object{LM~Vel}, and \object{Cyg~X-1} (from the HMXB catalog of \cite{Fortin2023}). The first one is in fact a runaway gamma-ray binary typically referred to as \object{LS~5039} \citep{Ribo2002}. In gamma-ray binaries the secondary mass transfer phase does not produce accretion onto the compact object but an intrabinary shock with the relativistic pulsar wind of the young NS \citep{Dubus2006,BoschRamon2012}. However, if the two objects in the SB1 system with a compact object are not close enough they would not interact, and these binaries could actually host quiescent stellar-mass BHs. \cite{Mahy2022} studied several SB1 systems with the aim of characterizing the nature of the unseen companions. They proposed our runaway SB1 system \object{HD~130~298} as a candidate to host a BH companion. Nevertheless, they also stated that confirming the BH nature of the companions requires further monitoring, sophisticated analysis techniques, and multiwavelength observations. \cite{Britavskiy2023} also proposed two of our runaway SB1 systems, \object{HD~94~024} and \object{HD~12~323}, as candidates for hosting a quiescent stellar-mass BH.

In Table~\ref{Tab:SB1_runaways}, we present an updated multiwavelength characterization of the 12 runaway stars classified here as SB1\footnote{We note that in a more detailed investigation of O-type stars \cite{Mahy2022} identified (through disentangling) a faint secondary in the spectra of \object{{HD~164 438}}, hence classifying the star as SB2.}, together with complementary information. For the systems in Table~\ref{Tab:SB1_runaways} with orbital parameters from \cite{Mahy2022}, we searched for correlations between $V_\text{PEC}^\text{2D}$ and $v \sin{i}$ as a function of $P_{\text{orb}}$ or $e$, but found no trends.

\renewcommand{\arraystretch}{1.1}
\begin{table*}
\centering
\tiny
\caption{Data of the runaway SB1 systems with the higher values of $E$ in decreasing order.}
\label{Tab:SB1_runaways}
\resizebox{\textwidth}{!}{\begin{tabular}{l@{~~~}r@{~}c@{~}l@{~~~}r@{~}c@{~}l@{~~~}c@{~~~}c@{~~~}c@{~~~}c@{~~~}r@{$~\pm~$}l@{~~~}c@{~~~}c@{~~~}c@{~~~}c@{~~~}c@{~~}c@{~~}c}
\hline\hline \vspace{-2mm}\\
GOSC Name & \multicolumn{3}{c}{RA} & \multicolumn{3}{c}{DEC} & $l$ & $b$ & $d$ & S.T. & \multicolumn{2}{c}{$V_\text{PEC}^\text{2D}$} & $E$ & $P_{\text{orb}}$ & $e$  & Radio & X-rays & Gamma rays & Comments \\
& (hh & mm &ss.ss) & ($\degr$ & $\arcmin$ & $\arcsec.\arcsec$) & ($\degr$) & ($\degr$) & (kpc) & & \multicolumn{2}{c}{(km~s$^{-1}$)} & & (d) & & \\
\hline \vspace{-2mm}\\
    \object{V479~Sct}    & 18 & 26 & 15.06   & $-$14 & 50 & 54.4   & ~~16.9  & $-$1.3  & 1.94   &  ON6V     & 90.9  & 4.9  & 4.28 & 3.91  & 0.254  & Detected\tablefootmark{a}   &Detected\tablefootmark{e} & HE\tablefootmark{j}\,VHE\tablefootmark{k}\,UHE\tablefootmark{l}& {HMXB}/$\gamma$-RB\\
    \object{HD~94~024}    & 10 & 50 & 01.50  & $-$57 & 52 & 26.2  & 287.3   & ~~1.3   &  2.59  & O8IV      & 49.9  & 4.3  & 2.49 & 2.46  & 0.000  &          &Not detected\tablefootmark{f} & &  cand BH       \\
    \object{HD~76~968}    & 08 & 57 & 28.85  & $-$50 & 44 & 58.2  & 270.2   & $-$3.4 &  2.21   & O9.2Ib    & 55.7  & 5.1  & 2.20 &       &        &         
 &Detected\tablefootmark{f} & &                \\
    \object{HD~46~573}    & 06 & 34 & 23.56  &  02   & 32 & 03.0  & 208.7   & $-$2.6 &  1.34   & O7V       & 40.4  & 4.6  & 1.98 & 10.65 & 0.595  &<151 $\mu$Jy\tablefootmark{b} &Not detected\tablefootmark{f} & &                \\
    \object{HD~130~298}   & 14 & 49 & 33.75  & $-$56 & 25 & 38.5  & 318.8   & ~~2.8  & 2.42    & O6.5III   & 30.4  & 3.2  & 1.76 & 14.63 & 0.457  &          &Not detected\tablefootmark{f} & &  cand BH       \\
    \object{HD~12~323}    & 02 & 02 & 30.12  &    55 & 37 & 26.3  & 132.9   & $-$5.9 & 2.27    & ON9.2V    & 53.4  & 9.5  & 1.68 & 1.93  & 0.000  &<187 $\mu$Jy\tablefootmark{b} &  & &  cand BH       \\
    \object{HDE~326~775}  & 17 & 05 & 31.31  & $-$41 & 31 & 20.2  & 345.0   & $-$0.3 &  1.58   & O6.5V     & 28.5  & 1.7  & 1.66 &       &        &<62 $\mu$Jy\tablefootmark{c} &Detected\tablefootmark{f} & &                \\
    \object{LM~Vel}      & 08 & 40 & 47.78  & $-$45 & 03 & 30.1  & 264.0   & $-$2.0 &  2.27   & O8.5Ib-II & 36.0  & 5.5  & 1.59 & 9.54  & 0.599  &           &Detected\tablefootmark{g,h} & &  HMXB             \\
    \object{HD~75~211}    & 08 & 47 & 01.58  & $-$44 & 04 & 28.7  & 264.0   & $-$0.5 &  1.59   &  O8.5II   & 23.6  & 3.0  & 1.15 & 20.45 & 0.340  &<110 $\mu$Jy\tablefootmark{c} &Detected\tablefootmark{f} & &                \\
    \object{Cyg~X-1}     & 19 & 58 & 21.67  &    35 & 12 & 05.7  & ~~71.3  & ~~3.1  &  2.16   & O9.7Iab   & 21.6  & 3.0  & 1.10 & 5.60  & 0.023  &Detected\tablefootmark{d} &Detected & HE\tablefootmark{m} &  HMXB             \\
    \object{HD~105~627}   & 12 & 09 & 44.57  & $-$62 & 34 & 54.6  & 298.2   & $-$0.1 &  2.21   & O9III     & 16.4  & 1.6  & 1.07 & 4.34  & 0.084  & <370 $\mu$Jy\tablefootmark{c}&Detected\tablefootmark{f,i} & &                \\
    \object{HD~164~438}   & 18 & 01 & 52.28  & $-$19 & 06 & 22.1  & ~~10.4  & ~~1.8  &  1.18   & O9.2IV    & 22.0  & 5.0  & 1.05 & 10.25 & 0.282  &     & &  &   MS comp\\
\hline
\end{tabular}}
\tablefoot{Columns up to $E$ are from \cite{MCC2023}, orbital period $P_{\text{orb}}$ and eccentricity $e$ are from \cite{Mahy2022} when available. Numerical values in the radio column correspond to the root mean square (RMS) of the noise level. Comments column contains additional information: `{HMXB}' for high-mass X-ray binaries \citep{Fortin2023}; `$\gamma$-RB' for gamma-ray binaries \citep{Bordas2024}. `cand BH' for systems with candidate BH companions \citep{Mahy2022,Britavskiy2023}. `MS comp' for Main Sequence companion \citep{Mahy2022}}.
\tablebib{Radio: $^{a}$\citet{Marti1998}; $^{b}$\citet{Lacy2020} (VLASS); $^{c}$\citet{Goedhart2024} (MeerKAT); $^{d}$\citet{Braes1971}. X-rays: $^{e}$\citet{Motch1997}; $^{f}$\citet{Freund2024} (eROSITA); $^{g}$\citet{Evans2014} (variable, Swift); $^{h}$\cite{Webb2023} (variable, XMM-Newton); $^{i}$\cite{Bowyer1965S}. Gamma rays: $^{j}$\citet{Paredes2000}; $^{k}$\citet{Aharonian2005}; $^{l}$\citet{Alfaro2025}; $^{m}$\citet{Zanin2016}.
}
\end{table*}

As already mentioned, there are three well-known systems that emit X-rays: \object{V479~Sct}/\object{LS~5039}, \object{LM~Vel}, and \object{Cyg~X-1}. In addition, there are four systems detected by {\it eROSITA}, but in all cases the X-ray emission has a high probability to be of coronal origin from the massive star
\citep{Freund2024}. \object{V479~Sct}/\object{LS~5039} and \object{Cyg~X-1} are well-known radio and high-energy (HE, $>100$~MeV) gamma-ray emitters. No other system shows either radio emission within VLASS or MeerKAT databases or HE emission within the {\it Fermi}-DR4 catalog \citep{Abdollahi2022,Ballet2023}. At very-high-energy (VHE, $>100$~GeV) gamma rays, \object{V479~Sct}/\object{LS~5039} has been clearly detected, while \object{Cyg~X-1} has been reported to show evidence of gamma-ray signal at 4-$\sigma$ significance \citep{Albert2007}. No other system is known to emit at VHEs. \object{V479~Sct}/\object{LS~5039} also shows ultra-high-energy (UHE, $>100$~TeV) gamma-ray emission. 

Moreover, an infrared (WISE W4 band, 12$\mu$m) search for stellar bow shocks among runaways conducted in \cite{MCC2025} shows that \object{HD~46~573} and \object{HD~130~298} display stellar bow shocks, whereas \object{LM~Vel}, \object{HD~75~211}, and \object{Cyg~X-1} show point-like emission, and \object{HD~164~438} a nearby curved doubtful IR structure.

\section{Discussion} \label{sec:discussion}

In this work we analyzed two samples: a sample of 214 Galactic O-type stars with available $v \sin{i}$ (rotation sample), and a subsample of those, comprising 168 targets, for which we have also available information about their LS/SB1 nature (rotation LS-SB1 subsample). We studied them in terms of $v \sin{i}$, $V_\text{PEC}^\text{2D}$, and binarity status. We found that: the majority of runaways and nearly all walkaways are slow rotators (items i and ii); that there are no fast-moving and fast-rotating runaways (iii); that the fraction of SB1 systems is always lower in runaways compared to normal stars (iv), with no runaway SB1 fast-rotating systems (v); and that in walkaways we find both LS and SB1 systems (vi).

In the following sections, we discuss these results in detail. We make a comparison with previous works in Sect.~\ref{sec:discussion_vsinivpec2d}, and interpret our results in terms of the runaway ejection scenarios in Sect.~\ref{sec:discussion_scenarios}. Finally, we discuss the findings of the runaway SB1 systems and comment on the runaway origin of the known HMXB in Sect.~\ref{sec:discussion_SB1}.

\subsection{Comparison with previous works} \label{sec:discussion_vsinivpec2d}

In Fig.~\ref{Fig:2D_rotsample} we presented the $v \sin{i}$ vs. $V_\text{PEC}^\text{2D}$ distribution for the 214 O runaway and normal stars in the rotation sample. We defined four regions within this figure (see Sect.~\ref{sec:results_2DrotS}). \citet{Sana2022} studied in a similar diagram a sample of 23 O-type runaway stars (among an investigated sample of 339 O-type stars) located in the 30 Dor region of the LMC. However, before comparing these two works, we should point out their differences. Beyond the different type of environment (star-forming region vs. a magnitude limited sample in the Milky Way\footnote{{In general, star-forming regions and clusters have higher stellar densities and more massive stars, which favor dynamical ejections of runaways into the field, where higher runaway fractions can be expected (e.g., \citealt{Dallas2022}). In contrast, volume-limited field samples may include runaways that have traveled far from their birthplaces. A detailed characterization of this effect is beyond the scope of this work.}}), they used 1D line-of-sight velocities to identify the runaway stars, while we used 2D peculiar velocities. They studied 23 O single runaway stars plus one runaway binary\footnote{This runaway binary was not studied within the four regions mentioned above.}, while our work includes 136 and 78 O normal and runaway stars, respectively, both single and binaries, and thus, our sample is significantly larger. Our runaway stars populate all four regions, although most are located in the slow-moving and slow-rotating one. This result is in contrast with the one of \cite{Sana2022}, who found that runaway stars mostly populate the slow-moving and fast-rotating region. They found 55\% of their runaways as fast rotators, while we only found 26\% within ours (see Table~\ref{Tab:rotSnormRun}). Thus, the investigated sample of O runaway stars in 30 Dor (LMC) and the one in our Galaxy may be dominated by different runaway ejection mechanisms (see Sect.~\ref{sec:discussion_scenarios}).

We have six fast-moving runaways with 2D peculiar velocities $\geq85$~km~s$^{-1}$. Their GOSC names (in order of increasing $V_\text{PEC}^\text{2D}$) are: 
\object{CPD~$-$34~2135}, 
\object{V479~Sct}, 
\object{HD~124~979},
\object{HD~157~857}, 
\object{HD~104~565}, and
\object{HD~116~852}.
All of them were identified in \cite{MCC2023}, but previously found by \cite{Ribo2002} and \cite{MA2018}. These six fast-moving runaways have 2D peculiar velocities from 85 to $\sim$220~km~s$^{-1}$ (although with large uncertainties at the high end), while \cite{Sana2022} have five fast-moving runaways with 1D peculiar velocities from 60 to $\sim$100~km~s$^{-1}$. Taking into account a factor of $\sqrt{2/1}$ between the 1D and 2D peculiar velocities, it seems that the runaway stars in the Galaxy can reach higher velocities than in 30 Dor within the LMC.

Finally, in the fast-moving and fast-rotating region \cite{Sana2022} did not find any runaway star and they qualify it as an avoidance region. However, we identified the runaway star \object{HD~124~979} within this avoidance region (with a relatively large uncertainty in  $V_\text{PEC}^\text{2D}$). This star was previously identified as SB2 by \cite{Sota2014}, but more recently also as single by \cite{Britavskiy2023}, as also found here. We comment more on the possible runaway origin of this object in the following section.

Although our work is focused on O-type stars, we note that \citet{Guo2024} also studied the rotational vs. space velocity distribution of Galactic early-type runaway stars, but in a sample clearly dominated by B-type stars (95\%). They found that most runaways are slow rotators with $v \sin{i}$ values up to only $\sim130$~km~s$^{-1}$. Nevertheless, we caution that they used a reduced fast-rotator threshold of $\sim60$~km~s$^{-1}$, much lower than ours of 200~km~$^{-1}$. Therefore, the four regions they examined within this plane are different from those studied here.

\subsection{Insights into the runaway ejection scenarios} \label{sec:discussion_scenarios}

\begin{figure*}
    \centering
    \includegraphics[width=0.8\hsize]{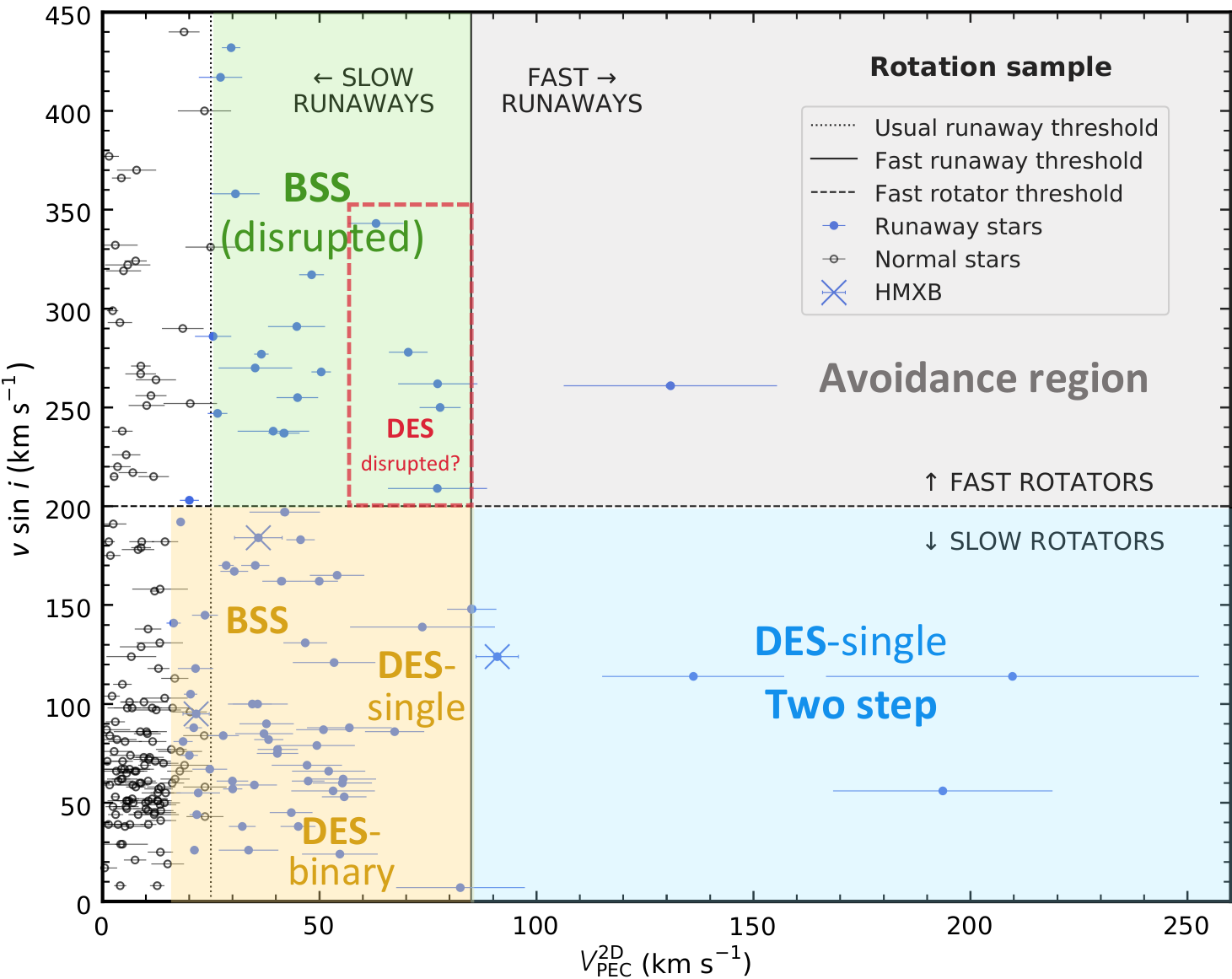}
    \caption{Interpretation of the possible runaway ejection scenarios in the $v \sin{i}$ vs. $V_\text{PEC}^\text{2D}$ plane. The labels related to the scenarios indicate the following. BSS: BSS ejection, resulting either in a bound or disrupted binary. BSS disrupted: BSS ejection of a former binary that becomes disrupted. DES disrupted: DES ejection of a former binary (note the high $v \sin{i}$) that becomes disrupted. DES-single: after DES ejection the runaway is identified as single. DES-binary: after DES ejection the runaway is identified as a binary. Two step: DES + BSS. Avoidance region: region with virtually no runaways. The colored regions are discussed in the text.}
    \label{Fig:scenarios}
\end{figure*}

The combination of $v \sin{i}$, $V_\text{PEC}^\text{2D}$, and binarity information to study runaway stars can shed light into their runaway origin. Here we discuss the possible interpretation of the runaway ejection scenarios across the different regions of the $v \sin{i}$ vs. $V_\text{PEC}^\text{2D}$ plane. We focus on the two primary scenarios, the BSS and DES, and comment on the possible combination of these mechanisms in the so-called two-step scenario. We provide a summary plot of our interpretation in Fig.~\ref{Fig:scenarios}. While this interpretation is qualitative rather than quantitative, we think it is useful to guide the reader through the discussion. It is also an attempt to provide a relation between different observational parameters and the runaway ejection scenarios.

The overall runaway percentage in the rotation sample is $\sim$36\%, which is higher than the $\sim$25\% found for the entire sample of GOSC-\textit{Gaia}~DR3 in \cite{MCC2023}. We note that when moving from the entire sample to the rotation sample we removed all SB2 systems. We computed the runaway percentage of the 117 LS, 51 SB1 and 65 SB2 systems identified in the GOSC-\textit{Gaia}~DR3 catalog with IACOB binarity data, and resulted in $\sim$45\%, $\sim$24\% and $\sim$10\%, respectively. Thus, SB2 systems have smaller runaway fractions. Simulations show that DES binary-binary interactions mostly result in the ejection of two single runaways and a binary (e.g., \citealt{Hoffer1983,Mikkola1983}). The binary, which is the most massive of the three products, is unlikely to gain a large velocity, and thus it is more difficult that SB2 systems acquire runaway velocities (e.g., \citealt{LeonardDuncan1988,Hoogerwerf2001}). Therefore, we attribute the higher runaway percentage found in the rotation sample to the absence of SB2 systems, which have low runaway fractions. In addition, as presented in Table~\ref{Tab:percent_rotLSSB1}, the fraction of SB1 systems is significantly higher among normal stars compared to runaway stars, 38\% and 18\%, respectively. Consequently, we note that as much as 82\% of the runaways appear to be single. Considering that the majority of massive stars are found in binaries (see \citealp{MarchantBodensteiner2024} for a review), these results contribute to the observational evidence that runaway ejection mechanisms in massive stars are in general associated with the disruption of binary systems. Nevertheless, we also found several SB1 runaway systems (see Sect.~\ref{sec:SB1runaways}). Thus, these runaway binary fractions can help to constrain the efficiency of disruption of binaries in runaway ejection simulations (e.g., \citealt{Eldridge2011,Renzo2019,Wagg2025}).

The fact that O-type fast-rotating stars are proposed as post-interaction binary products has important implications for understanding the origin of runaway stars. As explained in Sect.~\ref{Sec:Intro}, fast-rotating stars in post-interaction binaries, possibly become runaway stars via BSS after a SN explosion in the systems. As shown in Table~\ref{Tab:rotSslowFast}, we found a higher runaway fraction of 44\% for fast-rotating stars compared to 34\% for slow-rotating ones, supporting this connection. This result aligns with the one of \cite{Britavskiy2023}, who found a runaway fraction of $\sim$35--50\% among Galactic fast-rotating O-type stars compared to $\sim$20--30\% among slow-rotating ones. In addition, the median $v \sin{i}$ values for normal and runaway stars are $\sim$72 and $\sim$114~km~s$^{-1}$, respectively, indicating that runaway stars tend to rotate faster than normal stars, as already noted by \cite{MA2018}. This difference is statistically significant (see Sect.~\ref{sec:results_vsini_rotS}). Moreover, we did not identify fast-rotating runaway SB1 systems, even though such systems are present in the normal fast-rotating star domain, making identification bias an unlikely explanation (see Sect.~\ref{sec:results_rotLS_SB1}, and Tables~\ref{Tab:percent_rotLSSB1_regs}~and~\ref{Tab:percent_rotLSSB1}). Through binary population synthesis models, \cite{Renzo2019} predicted that most (86\%) BSS products are unbound after the SN explosion (though we caution that they obtained mostly walkaways, rather than runaways with $V_\text{PEC}^\text{3D}>30$~km~s$^{-1}$). This likely explains the absence of fast-rotating SB1 systems among runaway stars. Taking all this into consideration, these findings lead us to interpret the runaway stars in the fast-rotating domain as predominantly resulting from disrupted binaries through BSS (see upper-left green region in Fig.~\ref{Fig:scenarios}).

In contrast, BSS is not expected to operate as efficiently among slow rotators \citep{Hoogerwerf2001,MA2018,Britavskiy2023}. Therefore, the slow-rotating domain is more likely dominated by runaway stars produced through DES. Dynamical interactions in the DES mostly occur through single-binary or binary-binary encounters\footnote{While more complex multi-body interactions could exist, these ones are the most likely; thus, we simplify the discussion to these two cases for clarity.} \citep{Hoogerwerf2001,PeretsSubr2012}, resulting in different products such as single runaway stars and binary runaway systems, consistent with our results in binarity for slow-rotating runaways (see Fig.~\ref{Fig:2D_rotLS_SB1}). However, among the slow-rotating SB1 runaways we find the three known HMXBs discussed in Sect.~\ref{sec:SB1runaways}, {which could be explained due to tidal locking in relatively close binaries\footnote{In fact, the runaway SB1 systems presented in Table~\ref{Tab:SB1_runaways} have orbital periods smaller than 21 days, with typically $P_\text{orb}\lesssim10$ days.} (see \citealt{Hurley2002,Sen2022} for details)}. This indicates that the slow-rotating domain cannot be attributed exclusively to DES, since these binaries have at least experienced BSS\footnote{This is the case for NSs or BHs with SN explosion. BHs formed through direct collapse would require DES to explain their runaway nature.}. Thus, for the slow-moving and slow-rotating stars we propose a combination of DES (both single and binary interactions, with larger and lower velocities, respectively) and BSS (see bottom-left yellow region in Fig.~\ref{Fig:scenarios}). The relative contribution of each of them depends on factors such as the age and density of the parent clusters. Since the analysis in age or individual clusters is beyond the scope of this work, it is challenging to draw here quantitative conclusions for this regime.

Before discussing fast-moving and slow-rotating runaways, we examine those with $V_\text{PEC}^\text{2D}$ in the range 60--85~km~s$^{-1}$. The $V_\text{PEC}^\text{2D}$ distribution presents a drop around 60~km~s$^{-1}$ (see Fig.~\ref{Fig:2D_rotLS_SB1}-top). According to simulations, BSS disrupted products rarely reach velocities larger than 60~km~s$^{-1}$ \citep{Renzo2019}, and this is observationally found for HMXBs in the SMC \citep{Phillips2024}. Simulations also claim that large velocities are more likely obtained through DES rather than BSS, an interpretation already used in observational works \citep{Poveda1967,Leonard1991,PeretsSubr2012,DorigoJones2020,Phillips2024}. Therefore, we interpret runaways with velocities between 60 and 85~km~s$^{-1}$ as DES products. These runaways can be visually distinguished in Fig.~\ref{Fig:2D_rotLS_SB1}, both in the fast- and slow-rotating domains. Thus, we can speculate that the slow-moving and fast-rotating domain would also have a small contribution of DES (disrupted because the high $v \sin{i}$ suggests it was a former binary, see red discontinuous box in Fig.~\ref{Fig:scenarios}), while a contribution by DES-single could be present in the slow-moving and slow-rotating domain for $V_\text{PEC}^\text{2D}>60$~km~s$^{-1}$ (see the right part of the yellow region in Fig.~\ref{Fig:scenarios}).

Let us focus on the fast-moving and slow-rotating regime. Due to the conservation of linear momentum in dynamical interactions, the fastest runaway stars are expected to be single rather than binaries (e.g., \citealt{Leonard1991, PeretsSubr2012}). This is consistent with the results presented in Fig.~\ref{Fig:2D_rotLS_SB1}, where the fastest runaways are identified as LS, suggesting they may be DES-single products. In addition, the two-step scenario is another possibility to explain the fastest velocities of our runaways, as is probably the case of the gamma-ray binary \object{V479~Sct}/\object{LS~5039} with $V_\text{PEC}^\text{2D}\sim90$~km~s$^{-1}$ (see Sec.\ref{sec:discussion_SB1}). This hybrid scenario reconciles its unusually high velocity (unlikely via BSS alone) with its retained binarity. Therefore, we interpret the fast-moving and slow-rotating region as dominated by DES-single or the two-step scenario (bottom-right blue region in Fig.~\ref{Fig:scenarios}).

In the fast-moving and fast-rotating regime (top-right gray region in Fig.~\ref{Fig:scenarios}), we only found the runaway star \object{HD~124~979}, with no particular reference about its runaway origin in the literature. Given its $V_\text{PEC}^\text{2D}$ uncertainty, it could be eventually explained by the DES ejection of a former binary that becomes disrupted (DES disrupted). It is an interesting runaway star to be studied in more detail.

In any case, all runaway stars in Fig.~\ref{Fig:2D_rotsample} appear below the line $v \sin{i} = 500-1.75\,V_\text{PEC}^\text{2D}$~km~s$^{-1}$. This anticorrelation between maximum values of space velocity and rotational velocity is also seen in the green region of Fig.~\ref{Fig:scenarios}. 
We also note that all SB1 systems shown in Fig.~\ref{Fig:2D_rotLS_SB1} appear below the line $v \sin{i} = 320-2.4\,V_\text{PEC}^\text{2D}$~km~s$^{-1}$.  
These limits can be due to a combination of the different physical processes operating in the involved runaway ejection mechanisms. Further data with reduced uncertainties are needed to better constrain these observed trends and derive conclusions.

Regarding walkaway stars (left part of yellow region in Fig.~\ref{Fig:scenarios}), they have been theoretically proposed as slow-moving unbound BSS products. However, nearly all walkaway stars identified in our rotation sample are slow rotators, in contrast with BSS expectations due to a previous mass-transfer phase. In addition, within our rotation LS-SB1 subsample we found walkaways that are either likely single stars or binary systems. Therefore, not all of them are unbound. This suggests that walkaway stars cannot be exclusively interpreted as unbound BSS products. Instead, their origin likely involves a combination of BSS with a small kick velocity or DES-binary to explain the SB1 systems, and BSS with a large kick velocity or DES-single to explain the LS stars.

The different mechanisms and products of runaway binaries sketched in Fig.~\ref{Fig:scenarios} for fast and slow rotators, could eventually lead to statistical differences between these populations. The KS test presented in Fig.~\ref{Fig:CDF_vsini_rotS} found such differences when comparing the normal and runaway populations in $v \sin{i}$, particularly in the $\sim$50--250~km~s$^{-1}$ range. This points to a different origin of those distributions, which is expected due to the contribution of BSS in the fast rotating-domain. However, the KS tests presented in Appendix~\ref{sec:App_SpaceVelocities} and Sect.~\ref{sec:results_vpec2d} show that this is not the case in $V_\text{PEC}^\text{2D}$. Similarly, no significant differences were found when analyzing the $v \sin{i}$ in the context of binarity and runaway status (see Appendix~\ref{sec:App_CDF_rotbinS}). This could indicate either that a larger sample size is needed to detect such differences or that both the fast- and slow-rotating regimes are influenced by a combination of the BSS and DES that is difficult to disentangle.

We also compare our $v \sin{i}$ vs. $V_\text{PEC}^\text{2D}$ distribution with the theoretical one computed using binary population synthesis models and shown in Fig.~7 of \cite{Renzo2019}. We should first note three differences: their results are only for BSS at the time of ejection\footnote{Despite we work with the current velocity of the stars, differences caused by the different Galactic locations and Galactic potential on the trajectories are expected to be small \citep{MA2022a,Britavskiy2023}.}; they are for stars with $\geq7.5M_\sun$, while ours are for O-type stars with $\geq15M_\sun$; and they show the equatorial rotational velocity $v_{\text{eq}}$ while ours is the projected one. Despite these differences, we can still compare them in a qualitative way. The models predict a very pronounced peak in the fast-rotating domain at equatorial velocity of $\sim$500~km~s$^{-1}$, contrasting with our observed slow-rotation maximum around $\sim$60--100~km~s$^{-1}$ (see Fig.\ref{Fig:2D_rotsample}-right). Such a discrepancy would not be explained by projection effects \citep{Chandrasekhar1950}. Instead, their distribution appears more compatible with the 30 Dor results of \cite{Sana2022}, who concluded that their runaway population was dominated by BSS. In any case, we should interpret the simulations by \cite{Renzo2019} with caution, since they do not reproduce the generally observed runaway fractions and velocity distributions, and they point to several caveats, including the non-consideration of the DES contribution. Therefore, the difference between the results of the models and our observations may arise from an eventually relevant contribution of the DES in our sample of Galactic O-type runaway stars, as already note in \cite{MCC2023}, and as found for massive runaways OB stars in the SMC \citep{DorigoJones2020,Phillips2024}. 

Finally, \cite{Phillips2024} proposed that fast-rotating OB stars in the SMC do not appear to be dominated by BSS products, but might be runaway mergers produced by DES, as suggested by simulations in \cite{Leonard1995}. Their conclusion relies on the fact that they found statistically significant differences with a KS test between the $v \sin{i}$ distribution of all OBe stars with that of fast-rotating OB stars. However, this interpretation has limitations. First, the assumption that all OBe stars originate from BSS is overly simplistic (the authors also caution on this). Second, mergers are expected to rapidly spin down due to strong magnetic fields \citep{Ferrario2009,Schneider2019}, making them unlikely to remain as fast rotators. Third, only $\sim$1/3 of the mergers would acquire runaway velocities \citep{Leonard1995}, reducing the expected population. Finally, \cite{Leonard1995} claimed that merged runaways have higher multiplicity fractions, yet all fast-rotating runaways in our sample are single. Therefore, we still argue for the interpretation of runaway fast-rotating stars as predominantly produced by BSS.

\subsection{Runaway SB1 and HMXBs} \label{sec:discussion_SB1}

In Sect.~\ref{sec:SB1runaways} and Table~\ref{Tab:SB1_runaways} we presented the runaway SB1 systems identified in our sample. As explained there, runaway SB1 systems are candidates to be HMXBs. Indeed, this is the case for three out of 12 of the identified runaway SB1: \object{V479~Sct}/\object{LS~5039}, \object{LM~Vel}, and \object{Cyg~X-1}. In general, for a runaway HMXB to exist we need the formation of a compact object and an ejection mechanism. The compact object might be formed through a SN explosion or through direct BH collapse. Therefore, runaway HMXB can be formed through SN explosion within BSS, DES plus direct BH collapse, or the two-step scenario (DES + BSS).

\cite{Fortin2022} looked for the birthplaces of 26 Galactic HMXBs using \textit{Gaia}~EDR3 data, including our three runaways. For \object{V479~Sct}/\object{LS~5039}, they found no encounters with open clusters or spiral arms, even though it is located in a region of the Galactic plane with a higher cluster density than the average of their sample. They concluded that the isolation formation scenario (only BSS) is favored in the case of \object{V479~Sct}/\object{LS~5039}. However, taking into account that (1) \cite{Renzo2019} predicts that BSS-produced HMXBs should have systemic velocities of only a few tens of km~s$^{-1}$, and that (2) we found a large $V_\text{PEC}^\text{2D}$ of 91~km~s$^{-1}$, we propose that this system is a candidate for a two-step runaway ejection scenario. This could also explain why \cite{Fortin2022} did not find any direct encounter with a cluster for this object. For \object{LM~Vel}/\object{IGR~J08408$-$4503}, no encounters matching the expected age for the current massive star of the binary were identified by \cite{Fortin2022}. This lead the authors to conclude that it formed in isolation, consistent with a bound BSS. In the case of \object{Cyg~X-1}, its low $V_\text{PEC}^\text{2D}$ of 22~km~s$^{-1}$ suggests a BSS origin where the BH formed with minimal mass loss, or even through direct BH collapse as already noted by \cite{Mirabel2003}. While \cite{Fortin2022} found an encounter with the cluster \object{NGC~6871} (compatible with its age), the low $V_\text{PEC}^\text{2D}$ does not seem to be compatible with the DES scenario. Therefore, \object{Cyg~X-1} represents another bound BSS product. Both binary systems, \object{LM~Vel}/\object{IGR~J08408$-$4503} and \object{Cyg~X-1}, fall within the BSS (bound) region of Fig.~\ref{Fig:scenarios}, supporting this interpretation.

\begin{figure}
    \centering
    \includegraphics[width=\hsize]{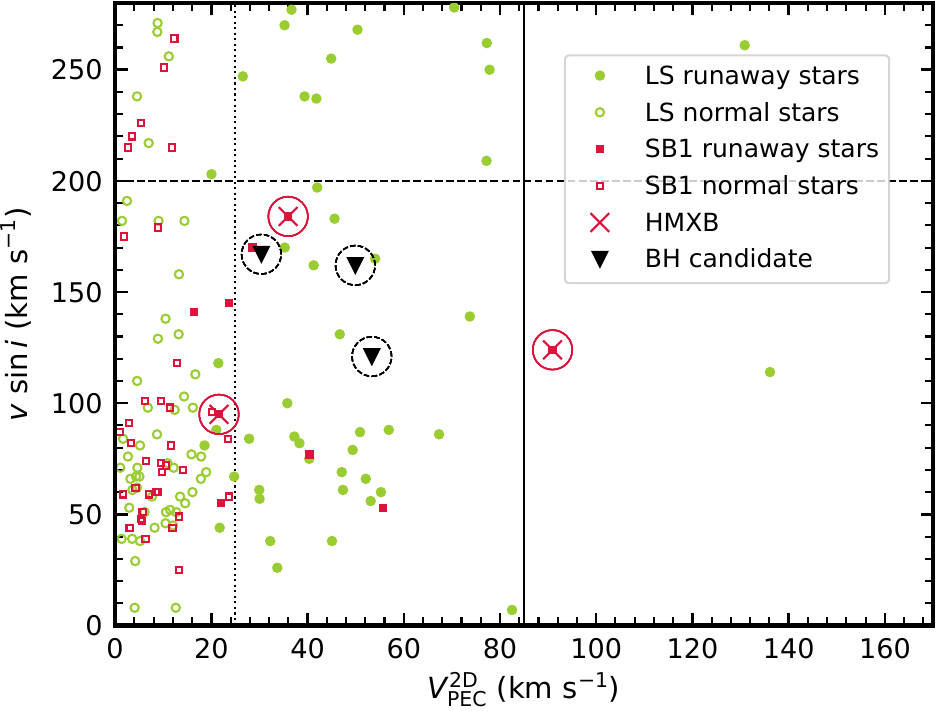}
    \caption{Lower-left part of the projected rotational velocity as a function of the 2D peculiar velocity plot presented in Fig.~\ref{Fig:2D_rotLS_SB1}. HMXBs are indicated with red crosses and circles, and binaries with BH candidates are indicated with black triangles and dashed circles.}
    \label{Fig:2D_rotLS_SB1_BHcand}
\end{figure}

The remaining nine runaway SB1 systems shown in Table~\ref{Tab:SB1_runaways} are interesting candidates to host compact objects, and potentially be HMXBs or gamma-ray binaries. In the case of HMXBs these could host NSs or BHs. While many HMXBs containing NSs are known, only a few of them host a confirmed BH \citep{CorralSantana2016}. Therefore, discovering new such systems would be very important to help constraining evolutionary models and to improve the statistics of BHs to better understand their formation scenarios. In particular, in Table~\ref{Tab:SB1_runaways} there are three binary systems which are candidates to host BHs: \object{HD~130~298} according to \cite{Mahy2022} and \object{HD~94~024} and \object{HD~12~323} according to \cite{Britavskiy2023}. Fig.~\ref{Fig:2D_rotLS_SB1_BHcand} shows the lower-left part of Fig.~\ref{Fig:2D_rotLS_SB1} including the locations of both HMXBs and BH candidates marked with red and dashed black circles, respectively. Interestingly, the HMXBs and the binaries with BH candidates occupy similar locations within the $v \sin{i}-V_\text{PEC}^\text{2D}$ parameter space, reinforcing their compact object candidancy. Therefore, these systems are particularly interesting to further study with dedicated multiwavelength observations. Even if hosting a stellar-mass BH, the expected accretion rates in these systems would be relatively low due to the orbital parameters and/or the spectral types of the massive companions (listed in Table~\ref{Tab:SB1_runaways}). However, their eventual detection at X-ray energies would point to low-luminosity analogs of \object{Cyg~X-1}, which could provide constraints on accretion processes, BH formation mechanisms, stellar evolution in binary systems and GW progenitors (e.g., \citealt{Sen2024}). In the case of gamma-ray binaries, only ten systems are known to exist at the time of writing \citep{Bordas2024}. Therefore, any addition to this small group would bring important information to help understanding these objects.

\section{Summary and conclusions} \label{Sec:Conclusions}

In this study, we conducted the largest analysis of kinematics $V_\text{PEC}^\text{2D}$, projected rotational velocities $v \sin{i}$, and binarity of Galactic O-type stars with accurate runaway classifications. Using \textit{Gaia}~DR3 astrometry and IACOB spectroscopy, we studied two samples: (1) an initial sample of 214 stars with $v \sin{i}$, and (2) a subsample of these comprising 168 targets with available information about their LS/SB1 nature. We computed fractions in the space of the mentioned parameters, statistically compared velocity distributions, and interpreted possible correlations between kinematics, rotation, binarity, and runaway ejection mechanisms. Below we summarize the main conclusions of this work.

\begin{itemize}

    \item Most Galactic O-type runaway and nearly all walkaway stars are slow rotators, {similar to what happens for normal stars. This is in} contrast to previous observational and theoretical works {of runaway stars}, pointing to a relevant contribution of DES.
    We find both LS and SB1 systems among slow rotators, with some HMXBs, indicating that a contribution of BSS is also present.

    \item The fraction of SB1 systems is always lower in runaways than in normal stars, implying that runaway ejection mechanisms typically disrupt binaries. The runaway percentages in LS, SB1, and SB2 systems are $\sim$45\%, $\sim$24\% and $\sim$10\%, respectively, revealing that single stars more likely obtained runaway velocities than binaries. SB2 systems present a low runaway fraction, hardly reaching runaway velocities, in agreement with DES simulations.

    \item The runaway fast rotators can be explained by the BSS because: there are no SB1 runaway fast-rotating systems, runaways rotate faster than normal stars on average, and there is significantly higher runaway fraction for fast-rotating than for slow-rotating stars. This is further supported by the result of the KS test between the normal and runaway star $v \sin{i}$ distributions, with statistical/significant differences mainly in the $50-250$~km~s$^{-1}$ range.

    \item Runaways with $V_\text{PEC}^\text{2D}>60$~km~s$^{-1}$ are mostly single and interpreted as DES products, while runaways with $V_\text{PEC}^\text{2D}>85$~km~s$^{-1}$ are also interpreted as two-step products, with \object{V479~Sct}/\object{LS~5039} a likely example.

    \item There are no fast-rotating and fast-moving O runaway stars, except the case of \object{HD~124~979}, which although it could be understood within DES, clearly deserves further attention.

    \item We found 12 runaway SB1 systems, 3 of which are HMXBs or gamma-ray binaries. The remaining systems are potential candidates to host compact objects as well, with 3 of them as candidates to host BHs that should be further studied.

\end{itemize}

We note that the results obtained in this work could be useful to constrain both, binary population synthesis models and simulations of stellar encounters within clusters. This would give us a better estimation of the efficiency of the different ejection scenarios. In addition, further studies could be conducted to compute the past trajectories of our runaways and relate them to parent clusters, which could help to put additional constraints on the scenarios. Finally, more extensive catalogs of massive stars and data from \textit{Gaia}-DR4 with improved astrometric uncertainties will add more and better statistics to test the observed trends in this work.

\begin{acknowledgements}

      {We thank the anonymous referee for useful comments. We thank J. Bodensteiner}, M. Gieles and M. Abdul-Masih for useful discussions. MC-C, MR, and JMP acknowledge financial support from the State Agency for Research of the Spanish Ministry of Science and Innovation under grants PID2022-136828NB-C41/AEI/10.13039/501100011033/ERDF/EU, and PID2022-138172NB-C43/AEI/10.13039/501100011033/ERDF/EU,   
      and through the Unit of Excellence María de Maeztu 2020-2023 and 2025-2029 awards to the Institute of Cosmos Sciences (CEX2019-000918-M, CEX2024-001451-M, MICIU/AEI/10.13039/501100011033). We acknowledge financial support from Departament de Recerca i Universitats of Generalitat de Catalunya through grant 2021SGR00679. MC-C acknowledges the grant PRE2020-094140 funded by MCIN/AEI/10.13039/501100011033 and FSE/ESF funds. GH, CMS, and SS-D acknowledge support from the Spanish Ministry of Science and Innovation and Universities (MICIU) through the Spanish State Research Agency (AEI) through grants PID2021-122397NB-C21, PID2022-136640NB-C22, 10.13039/501100011033, and the Severo Ochoa Program 2020-2023 (CEX2019-000920-S).
      This work has made use of data from the European Space Agency (ESA) mission {\it Gaia} (\url{https://www.cosmos.esa.int/gaia}), processed by the {\it Gaia} Data Processing and Analysis Consortium (DPAC, \url{https://www.cosmos.esa.int/web/gaia/dpac/consortium}). Funding for the DPAC has been provided by national institutions, in particular the institutions participating in the {\it Gaia} Multilateral Agreement. Regarding the observing facilities, this research is based on observations made with the Mercator Telescope, operated by the Flemish Community at the Observatorio del Roque de los Muchachos (La Palma, Spain), of the Instituto de Astrofísica de Canarias. In particular, obtained with the HERMES spectrograph, which is supported by the Research Foundation – Flanders (FWO), Belgium, the Research Council of KU Leuven, Belgium, the Fonds National de la Recherche Scientifique (F.R.S.-FNRS), Belgium, the Royal Observatory of Belgium, the Observatoire de Genève, Switzerland, and the Thüringer Landessternwarte Tautenburg, Germany. This work is also based on observations with the Nordic Optical Telescope, owned in collaboration by the University of Turku and Aarhus University, and operated jointly by Aarhus University, the University of Turku and the University of Oslo, representing Denmark, Finland and Norway, the University of Iceland and Stockholm University, at the Observatorio del Roque de los Muchachos, of the Instituto de Astrofísica de Canarias. Additionally this work is based on observations obtained with the FEROS spectrograph attached to the 2.2 m MPG/ESO telescope at the La Silla observatory (Chile). This research has made use of NASA’s Astrophysics Data System. This research has made use of the SIMBAD database, operated at CDS, Strasbourg, France.
\end{acknowledgements}

\bibliographystyle{aa}
\bibliography{mybibliography}

\begin{thebibliography}{117}
\expandafter\ifx\csname natexlab\endcsname\relax\def\natexlab#1{#1}\fi

\bibitem[{{Abdollahi} {et~al.}(2022){Abdollahi}, {Acero}, {Baldini}, {Ballet}, {Bastieri}, {Bellazzini}, {Berenji}, {Berretta}, {Bissaldi}, {Blandford}, {Bloom}, {Bonino}, {Brill}, {Britto}, {Bruel}, {Burnett}, {Buson}, {Cameron}, {Caputo}, {Caraveo}, {Castro}, {Chaty}, {Cheung}, {Chiaro}, {Cibrario}, {Ciprini}, {Coronado-Bl{\'a}zquez}, {Crnogorcevic}, {Cutini}, {D'Ammando}, {De Gaetano}, {Digel}, {Di Lalla}, {Dirirsa}, {Di Venere}, {Dom{\'\i}nguez}, {Fallah Ramazani}, {Fegan}, {Ferrara}, {Fiori}, {Fleischhack}, {Franckowiak}, {Fukazawa}, {Funk}, {Fusco}, {Galanti}, {Gammaldi}, {Gargano}, {Garrappa}, {Gasparrini}, {Giacchino}, {Giglietto}, {Giordano}, {Giroletti}, {Glanzman}, {Green}, {Grenier}, {Grondin}, {Guillemot}, {Guiriec}, {Gustafsson}, {Harding}, {Hays}, {Hewitt}, {Horan}, {Hou}, {J{\'o}hannesson}, {Karwin}, {Kayanoki}, {Kerr}, {Kuss}, {Landriu}, {Larsson}, {Latronico}, {Lemoine-Goumard}, {Li}, {Liodakis}, {Longo}, {Loparco}, {Lott}, {Lubrano}, {Maldera}, {Malyshev}, {Manfreda}, {Mart{\'\i}-Devesa},
  {Mazziotta}, {Mereu}, {Meyer}, {Michelson}, {Mirabal}, {Mitthumsiri}, {Mizuno}, {Moiseev}, {Monzani}, {Morselli}, {Moskalenko}, {Negro}, {Nuss}, {Omodei}, {Orienti}, {Orlando}, {Paneque}, {Pei}, {Perkins}, {Persic}, {Pesce-Rollins}, {Petrosian}, {Pillera}, {Poon}, {Porter}, {Principe}, {Rain{\`o}}, {Rando}, {Rani}, {Razzano}, {Razzaque}, {Reimer}, {Reimer}, {Reposeur}, {S{\'a}nchez-Conde}, {Saz Parkinson}, {Scotton}, {Serini}, {Sgr{\`o}}, {Siskind}, {Smith}, {Spandre}, {Spinelli}, {Sueoka}, {Suson}, {Tajima}, {Tak}, {Thayer}, {Thompson}, {Torres}, {Troja}, {Valverde}, {Wood}, \& {Zaharijas}}]{Abdollahi2022}
{Abdollahi}, S., {Acero}, F., {Baldini}, L., {et~al.} 2022, \apjs, 260, 53

\bibitem[{{Aharonian} {et~al.}(2005){Aharonian}, {Akhperjanian}, {Aye}, {Bazer-Bachi}, {Beilicke}, {Benbow}, {Berge}, {Berghaus}, {Bernl{\"o}hr}, {Boisson}, {Bolz}, {Borrel}, {Braun}, {Breitling}, {Brown}, {Gordo}, {Chadwick}, {Chounet}, {Cornils}, {Costamante}, {Degrange}, {Dickinson}, {Djannati-Ata{\"\i}}, {Drury}, {Dubus}, {Emmanoulopoulos}, {Espigat}, {Feinstein}, {Fleury}, {Fontaine}, {Fuchs}, {Funk}, {Gallant}, {Giebels}, {Gillessen}, {Glicenstein}, {Goret}, {Hadjichristidis}, {Hauser}, {Heinzelmann}, {Henri}, {Hermann}, {Hinton}, {Hofmann}, {Holleran}, {Horns}, {Jacholkowska}, {de Jager}, {Kh{\'e}lifi}, {Komin}, {Konopelko}, {Latham}, {Le Gallou}, {Lemi{\`e}re}, {Lemoine-Goumard}, {Leroy}, {Lohse}, {Marcowith}, {Martin}, {Martineau-Huynh}, {Masterson}, {McComb}, {de Naurois}, {Nolan}, {Noutsos}, {Orford}, {Osborne}, {Ouchrif}, {Panter}, {Pelletier}, {Pita}, {P{\"u}hlhofer}, {Punch}, {Raubenheimer}, {Raue}, {Raux}, {Rayner}, {Reimer}, {Reimer}, {Ripken}, {Rob}, {Rolland}, {Rowell}, {Sahakian},
  {Saug{\'e}}, {Schlenker}, {Schlickeiser}, {Schuster}, {Schwanke}, {Siewert}, {Sol}, {Spangler}, {Steenkamp}, {Stegmann}, {Tavernet}, {Terrier}, {Th{\'e}oret}, {Tluczykont}, {Vasileiadis}, {Venter}, {Vincent}, {V{\"o}lk}, \& {Wagner}}]{Aharonian2005}
{Aharonian}, F., {Akhperjanian}, A.~G., {Aye}, K.~M., {et~al.} 2005, Science, 309, 746

\bibitem[{{Albert} {et~al.}(2007){Albert}, {Aliu}, {Anderhub}, {Antoranz}, {Armada}, {Baixeras}, {Barrio}, {Bartko}, {Bastieri}, {Becker}, {Bednarek}, {Berger}, {Bigongiari}, {Biland}, {Bock}, {Bordas}, {Bosch-Ramon}, {Bretz}, {Britvitch}, {Camara}, {Carmona}, {Chilingarian}, {Coarasa}, {Commichau}, {Contreras}, {Cortina}, {Costado}, {Curtef}, {Danielyan}, {Dazzi}, {De Angelis}, {Delgado}, {de los Reyes}, {De Lotto}, {Domingo-Santamar{\'\i}a}, {Dorner}, {Doro}, {Errando}, {Fagiolini}, {Ferenc}, {Fern{\'a}ndez}, {Firpo}, {Flix}, {Fonseca}, {Font}, {Fuchs}, {Galante}, {Garc{\'\i}a-L{\'o}pez}, {Garczarczyk}, {Gaug}, {Giller}, {Goebel}, {Hakobyan}, {Hayashida}, {Hengstebeck}, {Herrero}, {H{\"o}hne}, {Hose}, {Hsu}, {Jacon}, {Jogler}, {Kosyra}, {Kranich}, {Kritzer}, {Laille}, {Lindfors}, {Lombardi}, {Longo}, {L{\'o}pez}, {L{\'o}pez}, {Lorenz}, {Majumdar}, {Maneva}, {Mannheim}, {Mansutti}, {Mariotti}, {Mart{\'\i}nez}, {Mazin}, {Merck}, {Meucci}, {Meyer}, {Miranda}, {Mirzoyan}, {Mizobuchi}, {Moralejo}, {Nieto},
  {Nilsson}, {Ninkovic}, {O{\~n}a-Wilhelmi}, {Otte}, {Oya}, {Panniello}, {Paoletti}, {Paredes}, {Pasanen}, {Pascoli}, {Pauss}, {Pegna}, {Persic}, {Peruzzo}, {Piccioli}, {Prandini}, {Puchades}, {Raymers}, {Rhode}, {Rib{\'o}}, {Rico}, {Rissi}, {Robert}, {R{\"u}gamer}, {Saggion}, {Saito}, {S{\'a}nchez}, {Sartori}, {Scalzotto}, {Scapin}, {Schmitt}, {Schweizer}, {Shayduk}, {Shinozaki}, {Shore}, {Sidro}, {Sillanp{\"a}{\"a}}, {Sobczynska}, {Stamerra}, {Stark}, {Takalo}, {Temnikov}, {Tescaro}, {Teshima}, {Torres}, {Turini}, {Vankov}, {Vitale}, {Wagner}, {Wibig}, {Wittek}, {Zandanel}, {Zanin}, \& {Zapatero}}]{Albert2007}
{Albert}, J., {Aliu}, E., {Anderhub}, H., {et~al.} 2007, \apjl, 665, L51

\bibitem[{{Alfaro} {et~al.}(2025){Alfaro}, {Araya}, \& {Arteaga-Vel\'azquez}}]{Alfaro2025}
{Alfaro}, R., {Araya}, M., \& {Arteaga-Vel\'azquez}, J.~C. 2025, arXiv e-prints, arXiv:2503.20947

\bibitem[{{Ballet} {et~al.}(2023){Ballet}, {Bruel}, {Burnett}, {Lott}, \& {The Fermi-LAT collaboration}}]{Ballet2023}
{Ballet}, J., {Bruel}, P., {Burnett}, T.~H., {Lott}, B., \& {The Fermi-LAT collaboration}. 2023, arXiv e-prints, arXiv:2307.12546

\bibitem[{{Bekenstein} \& {Bowers}(1974)}]{BekensteinBowers1974}
{Bekenstein}, J.~D. \& {Bowers}, R.~L. 1974, \apj, 190, 653

\bibitem[{{Blaauw}(1961)}]{Blaauw1961}
{Blaauw}, A. 1961, \bain, 15, 265

\bibitem[{{Blaauw}(1993)}]{Blaauw1993}
{Blaauw}, A. 1993, in Astronomical Society of the Pacific Conference Series, Vol.~35, Massive Stars: Their Lives in the Interstellar Medium, ed. J.~P. {Cassinelli} \& E.~B. {Churchwell}, 207

\bibitem[{{Bodensteiner} {et~al.}(2018){Bodensteiner}, {Baade}, {Greiner}, \& {Langer}}]{Bodensteiner2018}
{Bodensteiner}, J., {Baade}, D., {Greiner}, J., \& {Langer}, N. 2018, \aap, 618, A110

\bibitem[{{Bodensteiner} {et~al.}(2023){Bodensteiner}, {Sana}, {Dufton}, {Wang}, {Langer}, {Banyard}, {Mahy}, {de Koter}, {de Mink}, {Evans}, {G{\"o}tberg}, {H{\'e}nault-Brunet}, {Patrick}, \& {Schneider}}]{Bodensteiner2023}
{Bodensteiner}, J., {Sana}, H., {Dufton}, P.~L., {et~al.} 2023, \aap, 680, A32

\bibitem[{{Bordas}(2024)}]{Bordas2024}
{Bordas}, P. 2024, in 7th Heidelberg International Symposium on High-Energy Gamma-Ray Astronomy, 17

\bibitem[{{Bosch-Ramon} {et~al.}(2012){Bosch-Ramon}, {Barkov}, {Khangulyan}, \& {Perucho}}]{BoschRamon2012}
{Bosch-Ramon}, V., {Barkov}, M.~V., {Khangulyan}, D., \& {Perucho}, M. 2012, \aap, 544, A59

\bibitem[{{Bowyer} {et~al.}(1965){Bowyer}, {Byram}, {Chubb}, \& {Friedman}}]{Bowyer1965S}
{Bowyer}, S., {Byram}, E.~T., {Chubb}, T.~A., \& {Friedman}, H. 1965, Science, 147, 394

\bibitem[{{Braes} \& {Miley}(1971)}]{Braes1971}
{Braes}, L.~L.~E. \& {Miley}, G.~K. 1971, \nat, 232, 246

\bibitem[{{Britavskiy} {et~al.}(2023){Britavskiy}, {Sim{\'o}n-D{\'\i}az}, {Holgado}, {Burssens}, {Ma{\'\i}z Apell{\'a}niz}, {Eldridge}, {Naz{\'e}}, {Pantaleoni Gonz{\'a}lez}, \& {Herrero}}]{Britavskiy2023}
{Britavskiy}, N., {Sim{\'o}n-D{\'\i}az}, S., {Holgado}, G., {et~al.} 2023, \aap, 672, A22

\bibitem[{{Carretero-Castrillo} {et~al.}(2025){Carretero-Castrillo}, {Benaglia}, {Paredes}, \& {Rib{\'o}}}]{MCC2025}
{Carretero-Castrillo}, M., {Benaglia}, P., {Paredes}, J.~M., \& {Rib{\'o}}, M. 2025, \aap, 694, A250

\bibitem[{{Carretero-Castrillo} {et~al.}(2023){Carretero-Castrillo}, {Rib{\'o}}, \& {Paredes}}]{MCC2023}
{Carretero-Castrillo}, M., {Rib{\'o}}, M., \& {Paredes}, J.~M. 2023, \aap, 679, A109

\bibitem[{{Chandrasekhar} \& {M{\"u}nch}(1950)}]{Chandrasekhar1950}
{Chandrasekhar}, S. \& {M{\"u}nch}, G. 1950, \apj, 111, 142

\bibitem[{{Chini} {et~al.}(2012){Chini}, {Hoffmeister}, {Nasseri}, {Stahl}, \& {Zinnecker}}]{Chini2012}
{Chini}, R., {Hoffmeister}, V.~H., {Nasseri}, A., {Stahl}, O., \& {Zinnecker}, H. 2012, \mnras, 424, 1925

\bibitem[{{Conroy} \& {Kratter}(2012)}]{Conroy2012}
{Conroy}, C. \& {Kratter}, K.~M. 2012, \apj, 755, 123

\bibitem[{{Conti} \& {Ebbets}(1977)}]{ContiEbbets1977}
{Conti}, P.~S. \& {Ebbets}, D. 1977, \apj, 213, 438

\bibitem[{{Corral-Santana} {et~al.}(2016){Corral-Santana}, {Casares}, {Mu{\~n}oz-Darias}, {Bauer}, {Mart{\'\i}nez-Pais}, \& {Russell}}]{CorralSantana2016}
{Corral-Santana}, J.~M., {Casares}, J., {Mu{\~n}oz-Darias}, T., {et~al.} 2016, \aap, 587, A61

\bibitem[{{Dallas} {et~al.}(2022){Dallas}, {Oey}, \& {Castro}}]{Dallas2022}
{Dallas}, M.~M., {Oey}, M.~S., \& {Castro}, N. 2022, \apj, 936, 112

\bibitem[{{de Burgos} {et~al.}(2024{\natexlab{a}}){de Burgos}, {Keszthelyi}, {Sim{\'o}n-D{\'\i}az}, \& {Urbaneja}}]{deBurgos2024a}
{de Burgos}, A., {Keszthelyi}, Z., {Sim{\'o}n-D{\'\i}az}, S., \& {Urbaneja}, M.~A. 2024{\natexlab{a}}, \aap, 687, L16

\bibitem[{{de Burgos} {et~al.}(2025){de Burgos}, {Sim{\'o}n-D{\'\i}az}, {Urbaneja}, {Holgado}, {Ekstr{\"o}m}, {Ram{\'\i}rez-Tannus}, \& {Zari}}]{deBurgos2025}
{de Burgos}, A., {Sim{\'o}n-D{\'\i}az}, S., {Urbaneja}, M.~A., {et~al.} 2025, \aap, 695, A87

\bibitem[{{de Burgos} {et~al.}(2024{\natexlab{b}}){de Burgos}, {Sim{\'o}n-D{\'\i}az}, {Urbaneja}, \& {Puls}}]{deBurgos2024b}
{de Burgos}, A., {Sim{\'o}n-D{\'\i}az}, S., {Urbaneja}, M.~A., \& {Puls}, J. 2024{\natexlab{b}}, \aap, 687, A228

\bibitem[{{de Mink} {et~al.}(2012){de Mink}, {Brott}, {Cantiello}, {Izzard}, {Langer}, \& {Sana}}]{deMink2012}
{de Mink}, S.~E., {Brott}, I., {Cantiello}, M., {et~al.} 2012, in Astronomical Society of the Pacific Conference Series, Vol. 465, Proceedings of a Scientific Meeting in Honor of Anthony F. J. Moffat, ed. L.~{Drissen}, C.~{Robert}, N.~{St-Louis}, \& A.~F.~J. {Moffat}, 65

\bibitem[{{de Mink} {et~al.}(2009){de Mink}, {Cantiello}, {Langer}, {Pols}, {Brott}, \& {Yoon}}]{deMink2009}
{de Mink}, S.~E., {Cantiello}, M., {Langer}, N., {et~al.} 2009, \aap, 497, 243

\bibitem[{{de Mink} {et~al.}(2013){de Mink}, {Langer}, {Izzard}, {Sana}, \& {de Koter}}]{deMink2013}
{de Mink}, S.~E., {Langer}, N., {Izzard}, R.~G., {Sana}, H., \& {de Koter}, A. 2013, \apj, 764, 166

\bibitem[{{de Mink} {et~al.}(2014){de Mink}, {Sana}, {Langer}, {Izzard}, \& {Schneider}}]{deMink2014}
{de Mink}, S.~E., {Sana}, H., {Langer}, N., {Izzard}, R.~G., \& {Schneider}, F.~R.~N. 2014, \apj, 782, 7

\bibitem[{{Dervi{\c{s}}o{\v{g}}lu} {et~al.}(2010){Dervi{\c{s}}o{\v{g}}lu}, {Tout}, \& {Ibano{\v{g}}lu}}]{Dervisoglu2010}
{Dervi{\c{s}}o{\v{g}}lu}, A., {Tout}, C.~A., \& {Ibano{\v{g}}lu}, C. 2010, \mnras, 406, 1071

\bibitem[{{Dorigo Jones} {et~al.}(2020){Dorigo Jones}, {Oey}, {Paggeot}, {Castro}, \& {Moe}}]{DorigoJones2020}
{Dorigo Jones}, J., {Oey}, M.~S., {Paggeot}, K., {Castro}, N., \& {Moe}, M. 2020, \apj, 903, 43

\bibitem[{{Dubus}(2006)}]{Dubus2006}
{Dubus}, G. 2006, \aap, 456, 801

\bibitem[{{Eldridge} {et~al.}(2011){Eldridge}, {Langer}, \& {Tout}}]{Eldridge2011}
{Eldridge}, J.~J., {Langer}, N., \& {Tout}, C.~A. 2011, \mnras, 414, 3501

\bibitem[{{Evans} {et~al.}(2014){Evans}, {Osborne}, {Beardmore}, {Page}, {Willingale}, {Mountford}, {Pagani}, {Burrows}, {Kennea}, {Perri}, {Tagliaferri}, \& {Gehrels}}]{Evans2014}
{Evans}, P.~A., {Osborne}, J.~P., {Beardmore}, A.~P., {et~al.} 2014, \apjs, 210, 8

\bibitem[{{Ferrario} {et~al.}(2009){Ferrario}, {Pringle}, {Tout}, \& {Wickramasinghe}}]{Ferrario2009}
{Ferrario}, L., {Pringle}, J.~E., {Tout}, C.~A., \& {Wickramasinghe}, D.~T. 2009, \mnras, 400, L71

\bibitem[{{Fortin} {et~al.}(2022){Fortin}, {Garc{\'\i}a}, \& {Chaty}}]{Fortin2022}
{Fortin}, F., {Garc{\'\i}a}, F., \& {Chaty}, S. 2022, \aap, 665, A69

\bibitem[{{Fortin} {et~al.}(2023){Fortin}, {Garc{\'\i}a}, {Simaz Bunzel}, \& {Chaty}}]{Fortin2023}
{Fortin}, F., {Garc{\'\i}a}, F., {Simaz Bunzel}, A., \& {Chaty}, S. 2023, \aap, 671, A149

\bibitem[{{Freund} {et~al.}(2024){Freund}, {Czesla}, {Predehl}, {Robrade}, {Salvato}, {Schneider}, {Starck}, {Wolf}, \& {Schmitt}}]{Freund2024}
{Freund}, S., {Czesla}, S., {Predehl}, P., {et~al.} 2024, \aap, 684, A121

\bibitem[{Glivenko(1933)}]{Glivenko1933}
Glivenko, V. 1933, Gion. Ist. Ital. Attauri., 4, 92

\bibitem[{{Goedhart} {et~al.}(2024){Goedhart}, {Cotton}, {Camilo}, {Thompson}, {Umana}, {Bietenholz}, {Woudt}, {Anderson}, {Bordiu}, {Buckley}, {Buemi}, {Bufano}, {Cavallaro}, {Chen}, {Chibueze}, {Egbo}, {Frank}, {Hoare}, {Ingallinera}, {Irabor}, {Kraan-Korteweg}, {Kurapati}, {Leto}, {Loru}, {Mutale}, {Obonyo}, {Plavin}, {Rajohnson}, {Rigby}, {Riggi}, {Seidu}, {Serra}, {Smart}, {Stappers}, {Steyn}, {Surnis}, {Trigilio}, {Williams}, {Abbott}, {Adam}, {Asad}, {Baloyi}, {Bauermeister}, {Bennet}, {Bester}, {Botha}, {Brederode}, {Buchner}, {Burger}, {Cheetham}, {Cloete}, {de Villiers}, {de Villiers}, {du Toit}, {Esterhuyse}, {Fanaroff}, {Fourie}, {Gamatham}, {Gatsi}, {Geyer}, {Gouws}, {Gumede}, {Heywood}, {Hokwana}, {Hoosen}, {Horn}, {Horrell}, {Hugo}, {Isaacson}, {J{\'o}zsa}, {Jonas}, {Jordaan}, {Joubert}, {Julie}, {Kapp}, {Kriek}, {Kriel}, {Krishnan}, {Kusel}, {Legodi}, {Lehmensiek}, {Lord}, {Macfarlane}, {Magnus}, {Magozore}, {Main}, {Malan}, {Manley}, {Marais}, {Maree}, {Martens}, {Maruping}, {McAlpine},
  {Merry}, {Mgodeli}, {Millenaar}, {Mokone}, {Monama}, {New}, {Ngcebetsha}, {Ngoasheng}, {Nicolson}, {Ockards}, {Oozeer}, {Passmoor}, {Patel}, {Peens-Hough}, {Perkins}, {Ramaila}, {Ratcliffe}, {Renil}, {Richter}, {Salie}, {Sambu}, {Schollar}, {Schwardt}, {Schwartz}, {Serylak}, {Siebrits}, {Sirothia}, {Slabber}, {Smirnov}, {Tiplady}, {van Balla}, {van der Byl}, {Van Tonder}, {Venter}, {Venter}, {Welz}, \& {Williams}}]{Goedhart2024}
{Goedhart}, S., {Cotton}, W.~D., {Camilo}, F., {et~al.} 2024, \mnras, 531, 649

\bibitem[{{Guo} {et~al.}(2022){Guo}, {Liu}, {Wang}, {Wang}, {Zhang}, {Ji}, {Han}, \& {Chen}}]{Guo2022}
{Guo}, Y., {Liu}, C., {Wang}, L., {et~al.} 2022, \aap, 667, A44

\bibitem[{{Guo} {et~al.}(2024){Guo}, {Wang}, {Liu}, {Wu}, {Han}, \& {Chen}}]{Guo2024}
{Guo}, Y., {Wang}, L., {Liu}, C., {et~al.} 2024, \apjs, 272, 45

\bibitem[{{Hoffer}(1983)}]{Hoffer1983}
{Hoffer}, J.~B. 1983, \aj, 88, 1420

\bibitem[{{Holgado} {et~al.}(2018){Holgado}, {Sim{\'o}n-D{\'\i}az}, {Barb{\'a}}, {Puls}, {Herrero}, {Castro}, {Garcia}, {Ma{\'\i}z Apell{\'a}niz}, {Negueruela}, \& {Sab{\'\i}n-Sanjuli{\'a}n}}]{Holgado2018}
{Holgado}, G., {Sim{\'o}n-D{\'\i}az}, S., {Barb{\'a}}, R.~H., {et~al.} 2018, \aap, 613, A65

\bibitem[{{Holgado} {et~al.}(2020){Holgado}, {Sim{\'o}n-D{\'\i}az}, {Haemmerl{\'e}}, {Lennon}, {Barb{\'a}}, {Cervi{\~n}o}, {Castro}, {Herrero}, {Meynet}, \& {Arias}}]{Holgado2020}
{Holgado}, G., {Sim{\'o}n-D{\'\i}az}, S., {Haemmerl{\'e}}, L., {et~al.} 2020, \aap, 638, A157

\bibitem[{{Holgado} {et~al.}(2022){Holgado}, {Sim{\'o}n-D{\'\i}az}, {Herrero}, \& {Barb{\'a}}}]{Holgado2022}
{Holgado}, G., {Sim{\'o}n-D{\'\i}az}, S., {Herrero}, A., \& {Barb{\'a}}, R.~H. 2022, \aap, 665, A150

\bibitem[{{Hoogerwerf} {et~al.}(2001){Hoogerwerf}, {de Bruijne}, \& {de Zeeuw}}]{Hoogerwerf2001}
{Hoogerwerf}, R., {de Bruijne}, J.~H.~J., \& {de Zeeuw}, P.~T. 2001, \aap, 365, 49

\bibitem[{{Hurley} {et~al.}(2002){Hurley}, {Tout}, \& {Pols}}]{Hurley2002}
{Hurley}, J.~R., {Tout}, C.~A., \& {Pols}, O.~R. 2002, \mnras, 329, 897

\bibitem[{{Hut}(1981)}]{Hut1981}
{Hut}, P. 1981, \aap, 99, 126

\bibitem[{{Kimm} \& {Cen}(2014)}]{Kimm2014}
{Kimm}, T. \& {Cen}, R. 2014, \apj, 788, 121

\bibitem[{{Kobulnicky} \& {Chick}(2022)}]{Kobulnicky2022}
{Kobulnicky}, H.~A. \& {Chick}, W.~T. 2022, \aj, 164, 86

\bibitem[{{Kobulnicky} {et~al.}(2016){Kobulnicky}, {Chick}, {Schurhammer}, {Andrews}, {Povich}, {Munari}, {Olivier}, {Sorber}, {Wernke}, {Dale}, \& {Dixon}}]{Kobulnicky2016}
{Kobulnicky}, H.~A., {Chick}, W.~T., {Schurhammer}, D.~P., {et~al.} 2016, \apjs, 227, 18

\bibitem[{{Lacy} {et~al.}(2020){Lacy}, {Baum}, {Chandler}, {Chatterjee}, {Clarke}, {Deustua}, {English}, {Farnes}, {Gaensler}, {Gugliucci}, {Hallinan}, {Kent}, {Kimball}, {Law}, {Lazio}, {Marvil}, {Mao}, {Medlin}, {Mooley}, {Murphy}, {Myers}, {Osten}, {Richards}, {Rosolowsky}, {Rudnick}, {Schinzel}, {Sivakoff}, {Sjouwerman}, {Taylor}, {White}, {Wrobel}, {Andernach}, {Beasley}, {Berger}, {Bhatnager}, {Birkinshaw}, {Bower}, {Brandt}, {Brown}, {Burke-Spolaor}, {Butler}, {Comerford}, {Demorest}, {Fu}, {Giacintucci}, {Golap}, {G{\"u}th}, {Hales}, {Hiriart}, {Hodge}, {Horesh}, {Ivezi{\'c}}, {Jarvis}, {Kamble}, {Kassim}, {Liu}, {Loinard}, {Lyons}, {Masters}, {Mezcua}, {Moellenbrock}, {Mroczkowski}, {Nyland}, {O'Dea}, {O'Sullivan}, {Peters}, {Radford}, {Rao}, {Robnett}, {Salcido}, {Shen}, {Sobotka}, {Witz}, {Vaccari}, {van Weeren}, {Vargas}, {Williams}, \& {Yoon}}]{Lacy2020}
{Lacy}, M., {Baum}, S.~A., {Chandler}, C.~J., {et~al.} 2020, \pasp, 132, 035001

\bibitem[{{Langer} {et~al.}(2020){Langer}, {Sch{\"u}rmann}, {Stoll}, {Marchant}, {Lennon}, {Mahy}, {de Mink}, {Quast}, {Riedel}, {Sana}, {Schneider}, {Schootemeijer}, {Wang}, {Almeida}, {Bestenlehner}, {Bodensteiner}, {Castro}, {Clark}, {Crowther}, {Dufton}, {Evans}, {Fossati}, {Gr{\"a}fener}, {Grassitelli}, {Grin}, {Hastings}, {Herrero}, {de Koter}, {Menon}, {Patrick}, {Puls}, {Renzo}, {Sander}, {Schneider}, {Sen}, {Shenar}, {Sim{\'o}n-D{\'\i}as}, {Tauris}, {Tramper}, {Vink}, \& {Xu}}]{Langer2020}
{Langer}, N., {Sch{\"u}rmann}, C., {Stoll}, K., {et~al.} 2020, \aap, 638, A39

\bibitem[{{Larson}(1974)}]{Larson1974}
{Larson}, R.~B. 1974, \mnras, 169, 229

\bibitem[{{Leonard}(1991)}]{Leonard1991}
{Leonard}, P. J.~T. 1991, \aj, 101, 562

\bibitem[{{Leonard}(1995)}]{Leonard1995}
{Leonard}, P. J.~T. 1995, \mnras, 277, 1080

\bibitem[{{Leonard} \& {Duncan}(1988)}]{LeonardDuncan1988}
{Leonard}, P. J.~T. \& {Duncan}, M.~J. 1988, \aj, 96, 222

\bibitem[{{Lewin} \& {van der Klis}(2006)}]{Lewin2006}
{Lewin}, W. H.~G. \& {van der Klis}, M. 2006, {Compact Stellar X-ray Sources}, Vol.~39 (Cambridge University Press)

\bibitem[{{Mahy} {et~al.}(2022){Mahy}, {Sana}, {Shenar}, {Sen}, {Langer}, {Marchant}, {Abdul-Masih}, {Banyard}, {Bodensteiner}, {Bowman}, {Dsilva}, {Fabry}, {Hawcroft}, {Janssens}, {Van Reeth}, \& {Eldridge}}]{Mahy2022}
{Mahy}, L., {Sana}, H., {Shenar}, T., {et~al.} 2022, \aap, 664, A159

\bibitem[{{Ma{\'\i}z Apell{\'a}niz} {et~al.}(2024){Ma{\'\i}z Apell{\'a}niz}, {Negueruela}, \& {Caballero}}]{MA2024}
{Ma{\'\i}z Apell{\'a}niz}, J., {Negueruela}, I., \& {Caballero}, J.~A. 2024, arXiv e-prints, arXiv:2410.07301

\bibitem[{{Ma{\'\i}z Apell{\'a}niz} {et~al.}(2018){Ma{\'\i}z Apell{\'a}niz}, {Pantaleoni Gonz{\'a}lez}, {Barb{\'a}}, {Sim{\'o}n-D{\'\i}az}, {Negueruela}, {Lennon}, {Sota}, \& {Trigueros P{\'a}ez}}]{MA2018}
{Ma{\'\i}z Apell{\'a}niz}, J., {Pantaleoni Gonz{\'a}lez}, M., {Barb{\'a}}, R.~H., {et~al.} 2018, \aap, 616, A149

\bibitem[{{Ma{\'\i}z Apell{\'a}niz} {et~al.}(2022){Ma{\'\i}z Apell{\'a}niz}, {Pantaleoni Gonz{\'a}lez}, {Barb{\'a}}, \& {Weiler}}]{MA2022a}
{Ma{\'\i}z Apell{\'a}niz}, J., {Pantaleoni Gonz{\'a}lez}, M., {Barb{\'a}}, R.~H., \& {Weiler}, M. 2022, \aap, 657, A72

\bibitem[{{Ma{\'\i}z Apell{\'a}niz} {et~al.}(2013){Ma{\'\i}z Apell{\'a}niz}, {Sota}, {Morrell}, {Barb{\'a}}, {Walborn}, {Alfaro}, {Gamen}, {Arias}, \& {Gallego Calvente}}]{GOSC}
{Ma{\'\i}z Apell{\'a}niz}, J., {Sota}, A., {Morrell}, N.~I., {et~al.} 2013, in Massive Stars: From alpha to Omega, 198

\bibitem[{{Marchant} \& {Bodensteiner}(2024)}]{MarchantBodensteiner2024}
{Marchant}, P. \& {Bodensteiner}, J. 2024, \araa, 62, 21

\bibitem[{{Marcote} {et~al.}(2018){Marcote}, {Rib{\'o}}, {Paredes}, {Mao}, \& {Edwards}}]{Marcote2018}
{Marcote}, B., {Rib{\'o}}, M., {Paredes}, J.~M., {Mao}, M.~Y., \& {Edwards}, P.~G. 2018, \aap, 619, A26

\bibitem[{{Mart\'i} {et~al.}(1998){Mart\'i}, {Paredes}, \& {Rib\'o}}]{Marti1998}
{Mart\'i}, J., {Paredes}, J.~M., \& {Rib\'o}, M. 1998, \aap, 338, L71

\bibitem[{{Mart{\'\i}nez-Sebasti{\'a}n} {et~al.}(2025){Mart{\'\i}nez-Sebasti{\'a}n}, {Sim{\'o}n-D{\'\i}az}, {Jin}, {Keszthelyi}, {Holgado}, {Langer}, \& {Puls}}]{MartinezSebastian2025}
{Mart{\'\i}nez-Sebasti{\'a}n}, C., {Sim{\'o}n-D{\'\i}az}, S., {Jin}, H., {et~al.} 2025, \aap, 693, L10

\bibitem[{{Mikkola}(1983)}]{Mikkola1983}
{Mikkola}, S. 1983, \mnras, 203, 1107

\bibitem[{{Mirabel} \& {Rodrigues}(2003)}]{Mirabel2003}
{Mirabel}, I.~F. \& {Rodrigues}, I. 2003, Science, 300, 1119

\bibitem[{{Moe} \& {Di Stefano}(2017)}]{MoeStefano2017}
{Moe}, M. \& {Di Stefano}, R. 2017, \apjs, 230, 15

\bibitem[{{Mold{\'o}n} {et~al.}(2012){Mold{\'o}n}, {Rib{\'o}}, {Paredes}, {Brisken}, {Dhawan}, {Kramer}, {Lyne}, \& {Stappers}}]{Moldon2012}
{Mold{\'o}n}, J., {Rib{\'o}}, M., {Paredes}, J.~M., {et~al.} 2012, \aap, 543, A26

\bibitem[{{Motch} {et~al.}(1997){Motch}, {Haberl}, {Dennerl}, {Pakull}, \& {Janot-Pacheco}}]{Motch1997}
{Motch}, C., {Haberl}, F., {Dennerl}, K., {Pakull}, M., \& {Janot-Pacheco}, E. 1997, \aap, 323, 853

\bibitem[{{Naz{\'e}} {et~al.}(2023){Naz{\'e}}, {Britavskiy}, {Rauw}, {Labadie-Bartz}, \& {Sim{\'o}n-D{\'\i}az}}]{Naze2023}
{Naz{\'e}}, Y., {Britavskiy}, N., {Rauw}, G., {Labadie-Bartz}, J., \& {Sim{\'o}n-D{\'\i}az}, S. 2023, \mnras, 525, 1641

\bibitem[{{Negueruela} {et~al.}(2004){Negueruela}, {Steele}, \& {Bernabeu}}]{Negueruela2004}
{Negueruela}, I., {Steele}, I.~A., \& {Bernabeu}, G. 2004, Astronomische Nachrichten, 325, 749

\bibitem[{{Noriega-Crespo} {et~al.}(1997){Noriega-Crespo}, {van Buren}, \& {Dgani}}]{NoriegaCrespo1997}
{Noriega-Crespo}, A., {van Buren}, D., \& {Dgani}, R. 1997, \aj, 113, 780

\bibitem[{{Offner} {et~al.}(2023){Offner}, {Moe}, {Kratter}, {Sadavoy}, {Jensen}, \& {Tobin}}]{Offner2023}
{Offner}, S.~S.~R., {Moe}, M., {Kratter}, K.~M., {et~al.} 2023, in Astronomical Society of the Pacific Conference Series, Vol. 534, Protostars and Planets VII, ed. S.~{Inutsuka}, Y.~{Aikawa}, T.~{Muto}, K.~{Tomida}, \& M.~{Tamura}, 275

\bibitem[{{Oh} \& {Kroupa}(2016)}]{Oh_Kroupa_2016}
{Oh}, S. \& {Kroupa}, P. 2016, \aap, 590, A107

\bibitem[{{Packet}(1981)}]{Packet1981}
{Packet}, W. 1981, \aap, 102, 17

\bibitem[{{Paredes} {et~al.}(2000){Paredes}, {Mart{\'\i}}, {Rib{\'o}}, \& {Massi}}]{Paredes2000}
{Paredes}, J.~M., {Mart{\'\i}}, J., {Rib{\'o}}, M., \& {Massi}, M. 2000, Science, 288, 2340

\bibitem[{{Penny} \& {Gies}(2009)}]{PennyGies2009}
{Penny}, L.~R. \& {Gies}, D.~R. 2009, \apj, 700, 844

\bibitem[{{Perets} \& {{\v{S}}ubr}(2012)}]{PeretsSubr2012}
{Perets}, H.~B. \& {{\v{S}}ubr}, L. 2012, \apj, 751, 133

\bibitem[{{Peri} {et~al.}(2015){Peri}, {Benaglia}, \& {Isequilla}}]{Peri2015}
{Peri}, C.~S., {Benaglia}, P., \& {Isequilla}, N.~L. 2015, \aap, 578, A45

\bibitem[{{Petrovic} {et~al.}(2005){Petrovic}, {Langer}, \& {van der Hucht}}]{Petrovic2005}
{Petrovic}, J., {Langer}, N., \& {van der Hucht}, K.~A. 2005, \aap, 435, 1013

\bibitem[{{Pflamm-Altenburg} \& {Kroupa}(2010)}]{Pflamm-Altenburg2010}
{Pflamm-Altenburg}, J. \& {Kroupa}, P. 2010, \mnras, 404, 1564

\bibitem[{{Phillips} {et~al.}(2024){Phillips}, {Oey}, {Cuevas}, {Castro}, \& {Kothari}}]{Phillips2024}
{Phillips}, G.~D., {Oey}, M.~S., {Cuevas}, M., {Castro}, N., \& {Kothari}, R. 2024, \apj, 966, 243

\bibitem[{{Podsiadlowski} {et~al.}(1992){Podsiadlowski}, {Joss}, \& {Hsu}}]{Podsiadlowski1992}
{Podsiadlowski}, P., {Joss}, P.~C., \& {Hsu}, J.~J.~L. 1992, \apj, 391, 246

\bibitem[{{Pols} {et~al.}(1991){Pols}, {Cote}, {Waters}, \& {Heise}}]{Pols1991}
{Pols}, O.~R., {Cote}, J., {Waters}, L.~B.~F.~M., \& {Heise}, J. 1991, \aap, 241, 419

\bibitem[{{Portegies Zwart} {et~al.}(1999){Portegies Zwart}, {Makino}, {McMillan}, \& {Hut}}]{Portegies1999}
{Portegies Zwart}, S.~F., {Makino}, J., {McMillan}, S.~L.~W., \& {Hut}, P. 1999, \aap, 348, 117

\bibitem[{{Poveda} {et~al.}(1967){Poveda}, {Ruiz}, \& {Allen}}]{Poveda1967}
{Poveda}, A., {Ruiz}, J., \& {Allen}, C. 1967, Boletin de los Observatorios Tonantzintla y Tacubaya, 4, 86

\bibitem[{{Ram{\'\i}rez-Agudelo} {et~al.}(2015){Ram{\'\i}rez-Agudelo}, {Sana}, {de Mink}, {H{\'e}nault-Brunet}, {de Koter}, {Langer}, {Tramper}, {Gr{\"a}fener}, {Evans}, {Vink}, {Dufton}, \& {Taylor}}]{RamirezAgudelo2015}
{Ram{\'\i}rez-Agudelo}, O.~H., {Sana}, H., {de Mink}, S.~E., {et~al.} 2015, \aap, 580, A92

\bibitem[{{Ram{\'\i}rez-Agudelo} {et~al.}(2013){Ram{\'\i}rez-Agudelo}, {Sim{\'o}n-D{\'\i}az}, {Sana}, {de Koter}, {Sab{\'\i}n-Sanjul{\'\i}an}, {de Mink}, {Dufton}, {Gr{\"a}fener}, {Evans}, {Herrero}, {Langer}, {Lennon}, {Ma{\'\i}z Apell{\'a}niz}, {Markova}, {Najarro}, {Puls}, {Taylor}, \& {Vink}}]{RamirezAgudelo2013}
{Ram{\'\i}rez-Agudelo}, O.~H., {Sim{\'o}n-D{\'\i}az}, S., {Sana}, H., {et~al.} 2013, \aap, 560, A29

\bibitem[{{Renzo} {et~al.}(2019){Renzo}, {Zapartas}, {de Mink}, {G{\"o}tberg}, {Justham}, {Farmer}, {Izzard}, {Toonen}, \& {Sana}}]{Renzo2019}
{Renzo}, M., {Zapartas}, E., {de Mink}, S.~E., {et~al.} 2019, \aap, 624, A66

\bibitem[{{Rib{\'o}} {et~al.}(2002){Rib{\'o}}, {Paredes}, {Romero}, {Benaglia}, {Mart{\'\i}}, {Fors}, \& {Garc{\'\i}a-S{\'a}nchez}}]{Ribo2002}
{Rib{\'o}}, M., {Paredes}, J.~M., {Romero}, G.~E., {et~al.} 2002, \aap, 384, 954

\bibitem[{{Sana} {et~al.}(2012){Sana}, {de Mink}, {de Koter}, {Langer}, {Evans}, {Gieles}, {Gosset}, {Izzard}, {Le Bouquin}, \& {Schneider}}]{Sana2012}
{Sana}, H., {de Mink}, S.~E., {de Koter}, A., {et~al.} 2012, Science, 337, 444

\bibitem[{{Sana} {et~al.}(2022){Sana}, {Ram{\'\i}rez-Agudelo}, {H{\'e}nault-Brunet}, {Mahy}, {Almeida}, {de Koter}, {Bestenlehner}, {Evans}, {Langer}, {Schneider}, {Crowther}, {de Mink}, {Herrero}, {Lennon}, {Gieles}, {Ma{\'\i}z Apell{\'a}niz}, {Renzo}, {Sabbi}, {van Loon}, \& {Vink}}]{Sana2022}
{Sana}, H., {Ram{\'\i}rez-Agudelo}, O.~H., {H{\'e}nault-Brunet}, V., {et~al.} 2022, \aap, 668, L5

\bibitem[{{Schneider} {et~al.}(2019){Schneider}, {Ohlmann}, {Podsiadlowski}, {R{\"o}pke}, {Balbus}, {Pakmor}, \& {Springel}}]{Schneider2019}
{Schneider}, F. R.~N., {Ohlmann}, S.~T., {Podsiadlowski}, P., {et~al.} 2019, \nat, 574, 211

\bibitem[{{Sen} {et~al.}(2024){Sen}, {El Mellah}, {Langer}, {Xu}, {Quast}, \& {Pauli}}]{Sen2024}
{Sen}, K., {El Mellah}, I., {Langer}, N., {et~al.} 2024, \aap, 690, A256

\bibitem[{{Sen} {et~al.}(2022){Sen}, {Langer}, {Marchant}, {Menon}, {de Mink}, {Schootemeijer}, {Sch{\"u}rmann}, {Mahy}, {Hastings}, {Nathaniel}, {Sana}, {Wang}, \& {Xu}}]{Sen2022}
{Sen}, K., {Langer}, N., {Marchant}, P., {et~al.} 2022, \aap, 659, A98

\bibitem[{{Sim{\'o}n-D{\'\i}az} {et~al.}(2024){Sim{\'o}n-D{\'\i}az}, {Britavskiy}, {Castro}, {Holgado}, \& {de Burgos}}]{SimonDiaz2024}
{Sim{\'o}n-D{\'\i}az}, S., {Britavskiy}, N., {Castro}, N., {Holgado}, G., \& {de Burgos}, A. 2024, arXiv e-prints, arXiv:2405.11209

\bibitem[{{Sim{\'o}n-D{\'\i}az} {et~al.}(2017){Sim{\'o}n-D{\'\i}az}, {Godart}, {Castro}, {Herrero}, {Aerts}, {Puls}, {Telting}, \& {Grassitelli}}]{SimonDiaz2017}
{Sim{\'o}n-D{\'\i}az}, S., {Godart}, M., {Castro}, N., {et~al.} 2017, \aap, 597, A22

\bibitem[{{Sim{\'o}n-D{\'\i}az} \& {Herrero}(2014)}]{SimonDiaz2014}
{Sim{\'o}n-D{\'\i}az}, S. \& {Herrero}, A. 2014, \aap, 562, A135

\bibitem[{{Sim{\'o}n-D{\'\i}az} {et~al.}(2020){Sim{\'o}n-D{\'\i}az}, {P{\'e}rez Prieto}, {Holgado}, {de Burgos}, \& {Iacob Team}}]{SimonDiaz2020}
{Sim{\'o}n-D{\'\i}az}, S., {P{\'e}rez Prieto}, J.~A., {Holgado}, G., {de Burgos}, A., \& {Iacob Team}. 2020, in XIV.0 Scientific Meeting (virtual) of the Spanish Astronomical Society, 187

\bibitem[{Smirnov(1939)}]{Smirnov1939}
Smirnov, N.~V. 1939, Bull. Math. Univ. Moscou, 2, 3

\bibitem[{{Sota} {et~al.}(2014){Sota}, {Ma{\'\i}z Apell{\'a}niz}, {Morrell}, {Barb{\'a}}, {Walborn}, {Gamen}, {Arias}, \& {Alfaro}}]{Sota2014}
{Sota}, A., {Ma{\'\i}z Apell{\'a}niz}, J., {Morrell}, N.~I., {et~al.} 2014, \apjs, 211, 10

\bibitem[{{Stone}(1979)}]{Stone1979}
{Stone}, R.~C. 1979, \apj, 232, 520

\bibitem[{{Stone}(1982)}]{Stone1982}
{Stone}, R.~C. 1982, \apj, 261, 208

\bibitem[{{Stoop} {et~al.}(2024){Stoop}, {de Koter}, {Kaper}, {Brands}, {Portegies Zwart}, {Sana}, {Stoppa}, {Gieles}, {Mahy}, {Shenar}, {Guo}, {Nelemans}, \& {Rieder}}]{Stoop2024}
{Stoop}, M., {de Koter}, A., {Kaper}, L., {et~al.} 2024, \nat, 634, 809

\bibitem[{{Tetzlaff} {et~al.}(2011){Tetzlaff}, {Neuh{\"a}user}, \& {Hohle}}]{Tetzlaff2011}
{Tetzlaff}, N., {Neuh{\"a}user}, R., \& {Hohle}, M.~M. 2011, \mnras, 410, 190

\bibitem[{{Tylenda} {et~al.}(2011){Tylenda}, {Hajduk}, {Kami{\'n}ski}, {Udalski}, {Soszy{\'n}ski}, {Szyma{\'n}ski}, {Kubiak}, {Pietrzy{\'n}ski}, {Poleski}, {Wyrzykowski}, \& {Ulaczyk}}]{Tylenda2011}
{Tylenda}, R., {Hajduk}, M., {Kami{\'n}ski}, T., {et~al.} 2011, \aap, 528, A114

\bibitem[{{van den Heuvel}(2007)}]{Heuvel2007}
{van den Heuvel}, E.~P.~J. 2007, in American Institute of Physics Conference Series, Vol. 924, The Multicolored Landscape of Compact Objects and Their Explosive Origins, ed. T.~{di Salvo}, G.~L. {Israel}, L.~{Piersant}, L.~{Burderi}, G.~{Matt}, A.~{Tornambe}, \& M.~T. {Menna}, 598--606

\bibitem[{{van Oijen}(1989)}]{Oijen1989}
{van Oijen}, J.~G.~J. 1989, \aap, 217, 115

\bibitem[{{Wagg} {et~al.}(2025){Wagg}, {Breivik}, {Renzo}, \& {Price-Whelan}}]{Wagg2025}
{Wagg}, T., {Breivik}, K., {Renzo}, M., \& {Price-Whelan}, A.~M. 2025, \apjs, 276, 16

\bibitem[{{Webb} {et~al.}(2023){Webb}, {Coriat}, {Traulsen}, {Ballet}, {Motch}, {Carrera}, {Koliopanos}, {Authier}, {de La Calle}, {Ceballos}, {Colomo}, {Chuard}, {Freyberg}, {Garcia}, {Kolehmainen}, {Lamer}, {Lin}, {Maggi}, {Michel}, {Page}, {Page}, {Perea-Calderon}, {Pineau}, {Rodriguez}, {Rosen}, {Santos Lleo}, {Saxton}, {Schwope}, {Tomas}, {Watson}, \& {Zakardjian}}]{Webb2023}
{Webb}, N.~A., {Coriat}, M., {Traulsen}, I., {et~al.} 2023, {VizieR Online Data Catalog: XMM-Newton Serendipitous Source Catalogue 4XMM-DR13 (Webb+, 2023)}, VizieR On-line Data Catalog: IX/69. Originally published in: 2020A\&A...641A.136W

\bibitem[{{Zahn}(1975)}]{Zahn1975}
{Zahn}, J.~P. 1975, \aap, 41, 329

\bibitem[{{Zanin} {et~al.}(2016){Zanin}, {Fern{\'a}ndez-Barral}, {de O{\~n}a Wilhelmi}, {Aharonian}, {Blanch}, {Bosch-Ramon}, \& {Galindo}}]{Zanin2016}
{Zanin}, R., {Fern{\'a}ndez-Barral}, A., {de O{\~n}a Wilhelmi}, E., {et~al.} 2016, \aap, 596, A55

\end{thebibliography}

\begin{appendix}

\onecolumn
\section{Data of the rotation sample} \label{sec:App_rotsample}

\renewcommand{\arraystretch}{1.1}
\begin{table*}[h]
\centering
\tiny
\caption{Data of the first 10 stars of the rotation sample with the higher values of $E$ in decreasing order.}
\label{Tab:rotsample}
\resizebox{\textwidth}{!}{\begin{tabular}{l@{~~~}c@{~~~}r@{~}c@{~}l@{~~~}r@{~}c@{~}l@{~~~}c@{~~~}c@{~~~}c@{~~~}c@{~~~}c@{~~~}r@{$~\pm~$}l@{~~~}c@{~~~}c@{~~~}}
\hline\hline \vspace{-2mm}\\
GOSC Name & \textit{Gaia} DR3 id & \multicolumn{3}{c}{RA} & \multicolumn{3}{c}{DEC} & $l$ & $b$ & $d$ & $G$ & S.T. & \multicolumn{2}{c}{$V_\text{PEC}^\text{2D}$} & $E$ & $v \sin{i}$ \\
& & (hh & mm &ss.ss) & ($\degr$ & $\arcmin$ & $\arcsec.\arcsec$) & ($\degr$) & ($\degr$) & (kpc) & & & \multicolumn{2}{c}{(km~s$^{-1}$)} &      & (km~s$^{-1}$)\\
\hline \vspace{-2mm}\\
  \object{V479~Sct}       & 4104196427943626624 & 18 & 26 & 15.06 & $-$14 & 50 & 54.4 & ~~16.9 & $-$1.3 & 1.94 &  10.80 & ON6   & 90.9  & 4.9    & 4.28  & 124\\
  \object{CPD~$-$34~2135} & 5546501254035205376 & 08 & 13 & 35.36 & $-$34 & 28 & 43.9 &  252.4 & $-$0.1 & 3.23 & ~~9.18 & O7.5  & 85.1  & 5.6    & 3.85  & 148\\
  \object{BD~$+$60~134}   & 427457895747434880  & 00 & 56 & 14.21 &    61 & 45 & 36.9 &  123.5 & $-$1.1 & 2.69 &  10.40 & O5.5  & 77.9  & 4.8    & 3.70  & 250 \\
  \object{HD~104~565}     & 6072058878595295488 & 12 & 02 & 27.77 & $-$58 & 14 & 34.3 &  296.5 &  ~~4.0 & 4.72 & ~~9.07 & OC9.7 & 193.7 & 25.2   & 3.47  & 56\\
  \object{HD~155~913}     & 5953699131931631232 & 17 & 16 & 26.32 & $-$42 & 40 & 04.0 &  345.3 & $-$2.6 & 1.21 & ~~8.18 & O4.5  & 70.6  & 4.5    & 3.43  & 278\\
  \object{HD~75~222}      & 5625488726258364544 & 08 & 47 & 25.13 & $-$36 & 45 & 02.5 &  258.3 &  ~~4.2 & 2.09 & ~~7.31 & O9.7  & 67.4  & 6.9    & 3.03  & 86 \\  
  \object{BD~$-$14~5040}  & 4104201586232296960 & 18 & 25 & 38.91 & $-$14 & 45 & 05.8 & ~~16.9 & $-$1.1 & 1.63 &  10.02 & O5.5  & 50.4  & 2.3    & 2.87  & 268\\
  \object{HD~41~997}      & 3345950879898371712 & 06 & 08 & 55.82 &    15 & 42 & 18.0 &  194.2 & $-$2.0 & 1.75 & ~~8.30 & O7.5  & 77.3  & 9.1    & 2.79  & 262\\
  \object{HD~157~857}     & 4165963456443515520 & 17 & 26 & 17.32 & $-$10 & 59 & 34.8 &  ~~13.0  &   13.3 & 2.30 & ~~7.71 & O6.5  & 136.1 & 21.0   & 2.60  & 114\\
  \object{HD~94~024}      & 5350908374676231808 & 10 & 50 & 01.50 & $-$57 & 52 & 26.2 &  287.3 &  ~~1.3 & 2.59 & ~~8.68 & O8    & 49.9  & 4.3    & 2.49  & 162\\
  
\hline
\end{tabular}}
\tablefoot{Columns up to $E$ were computed in \cite{MCC2023}. Most of the $v \sin{i}$ values are from \cite{Holgado2022}. The complete version of this table, including the 214 normal and runaway stars with additional parameters and precision, is available at the CDS.}
\end{table*}

\FloatBarrier
~
\newpage


\twocolumn
\section{2D space velocity distributions}\label{sec:App_SpaceVelocities}

\begin{figure}[h!!!]
    \centering
    \includegraphics[width=\hsize]{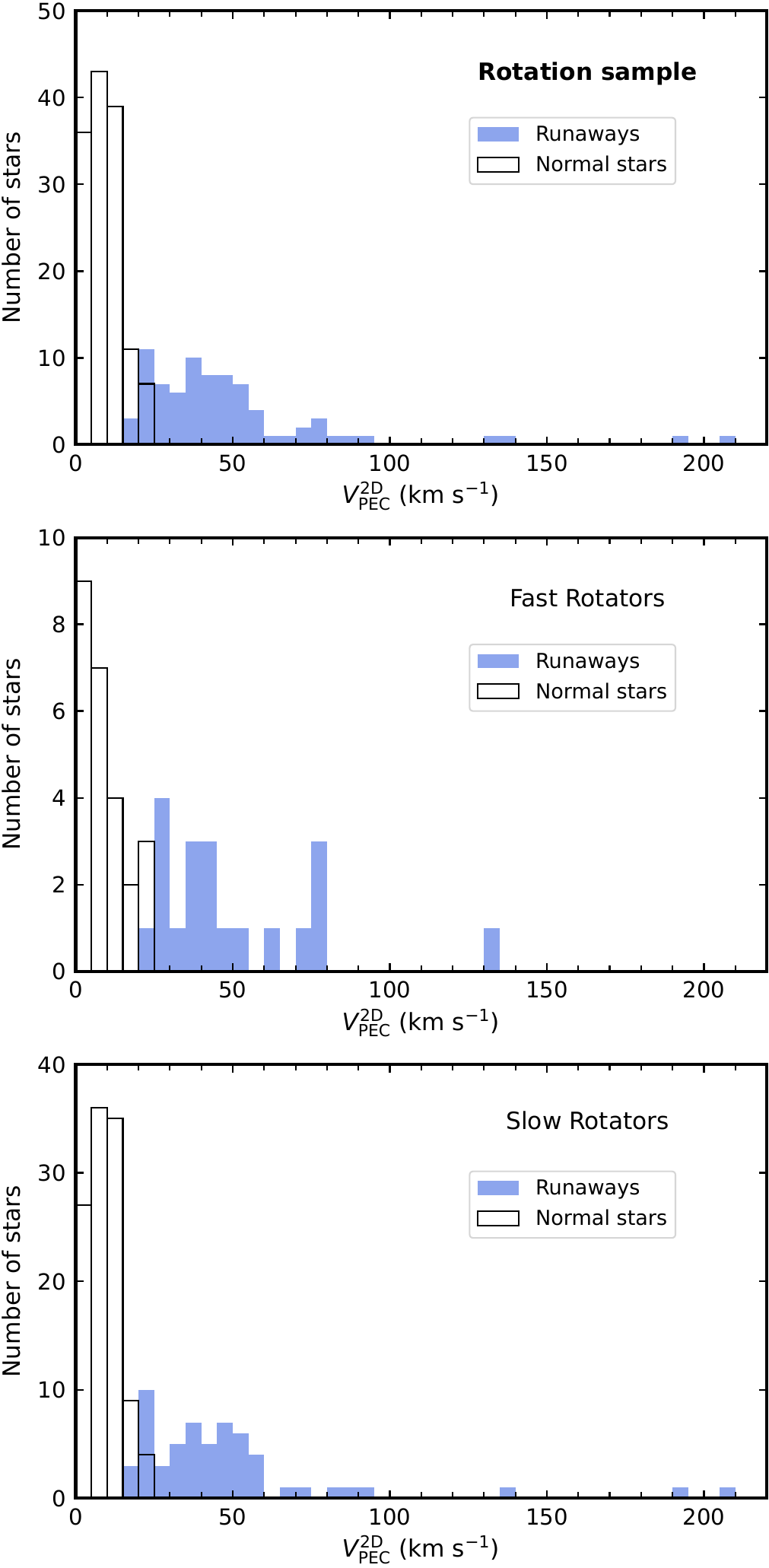}
    \caption{$V_\text{PEC}^\text{2D}$ distributions for the normal stars (unfilled black bars) and runaway stars (filled blue bars) with a bin size of 5~km~s$^{-1}$, for the rotation sample (top), the fast rotators (middle), and the slow rotators (bottom).}
    \label{Fig:vpec2d_run_norm}
\end{figure}

We present in Fig.~\ref{Fig:vpec2d_run_norm} the $V_\text{PEC}^\text{2D}$ distribution for the entire rotation sample (top), the fast (middle) and the slow (bottom) rotators, subdividing them into normal and runaway stars. The distribution of the slow-rotating runaway stars resembles the one of the full rotation sample, while the distribution of the fast-rotating runaways seems compatible with a flat one considering the poor statistics. The medians and 95th-percentiles of these distributions are presented in Sect.~\ref{sec:results_vpec2d}.

\begin{figure*}[t!]
    \centering
    \includegraphics[width=\hsize]{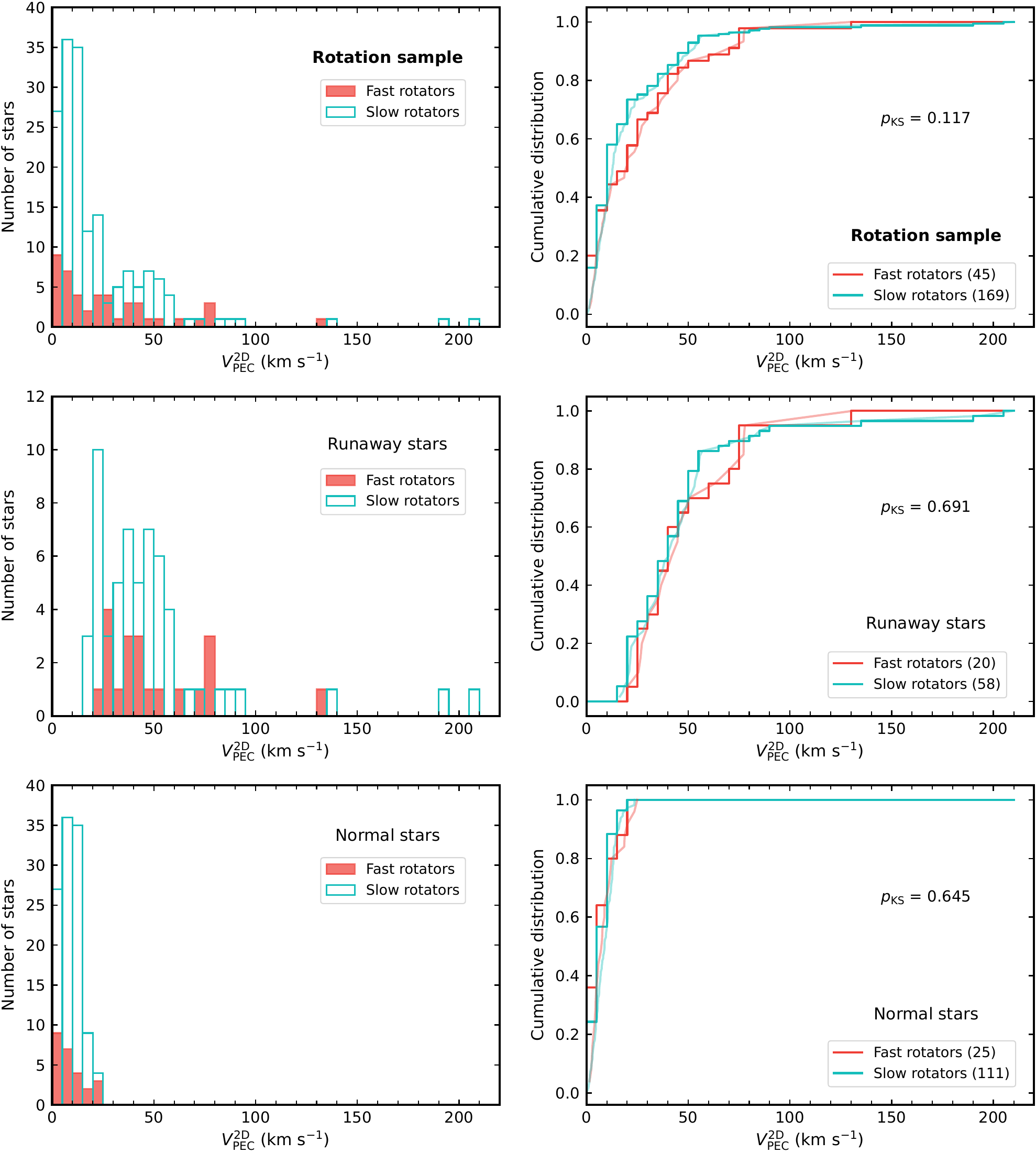}
    \caption{\textit{Left}: $V_\text{PEC}^\text{2D}$ distributions for the slow-rotating (unfilled turquoise bars) and fast-rotating (filled pale red bars) stars with a bin size of 5~km~s$^{-1}$ for the rotation sample (top), the runaway stars (middle) and normal stars (bottom). \textit{Right}: Cumulative distribution functions of $V_\text{PEC}^\text{2D}$ for the slow- (turquoise) and fast-rotating (pale red) stars for the rotation sample (top), the runaway stars (middle), and the normal stars (bottom). The curves in light colors show the corresponding empirical cumulative distribution functions. The numbers of fast and slow rotators are shown between parentheses. $p_{\rm KS}$ are the corresponding $p$-values resulting from the KS tests.}
      \label{Fig:vpec2d_slow_fast}
\end{figure*}

Instead, in the left panels of Fig.~\ref{Fig:vpec2d_slow_fast} we present the $V_\text{PEC}^\text{2D}$ distributions for the entire rotation sample (top), the runaway (middle) and the normal (bottom) stars, subdividing them into fast and slow rotators. In the right panels of the figure we also present the CDFs of $V_\text{PEC}^\text{2D}$ for the same three sets. The CDFs for the slow- and fast-rotating runaway stars appear somehow different. To check if these differences are statistically significant, we used the Kolmogorov-Smirnov (KS) test \citep{Smirnov1939}, which allows us to test the null hypothesis that two observed distributions are randomly drawn from the same parent population. The results of the KS tests between the fast and the slow-rotating stars for each of the three sets are included within the panels of the presented CDFs (right-panels of Fig.~\ref{Fig:vpec2d_slow_fast}). Although the KS test between the fast and slow-rotating sets in the rotation sample (top panel) may indicate that the distributions could be different, the results are not statistically significant for any of the rotation sample, runaway, and normal star sets. In particular, we note that in the case of the runaway stars, this may be due to the poor statistics for the fast-rotating stars, which are only 20.

\FloatBarrier
~
\newpage
\section{Cumulative distributions considering information on binarity} \label{sec:App_CDF_rotbinS}

\begin{figure}[h!!!]
    \centering
    \includegraphics[width=\hsize]{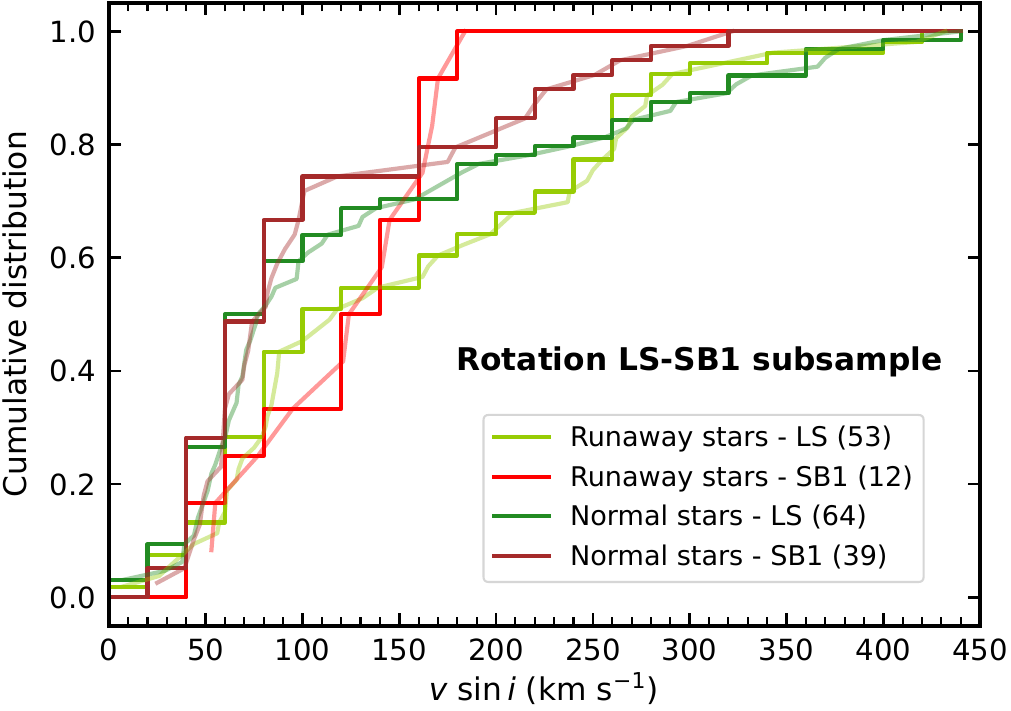}
    \caption{Cumulative distribution functions of $v \sin{i}$ for the different sets of the rotation LS-SB1 subsample. LS stars are illustrated in dark and light green, for the normal and runaway stars, respectively. SB1 systems are illustrated in dark and light red, for the normal and runaway stars, respectively. The curves in light colors show the corresponding empirical cumulative distribution functions.}
    \label{Fig:CDFvsini_LS_SB1}
\end{figure}

We present in Fig.~\ref{Fig:CDFvsini_LS_SB1} the CDFs of $v \sin{i}$ for four different sets: runaway and normal stars, each subdivided into SB1 and LS categories. The corresponding KS test results are summarized in Table~\ref{Tab:KStest_LS_SB1}. While there are no statistically significant differences between these populations, we note that some comparisons between normal and runaway stars, for both LS and SB1 categories, result in $p_{\text{KS}}\sim7-9$\%, which approaches the conventional 5\% significance threshold.

\FloatBarrier

\renewcommand{\arraystretch}{1.1}
\begin{table}[h!]
    \centering
    \caption{Results of the Kolmogorov-Smirnov tests in $v \sin{i}$ for different sets: Runaway LS stars (Run-LS), runaway SB1 systems (Run-SB1), normal LS stars (Norm-LS), normal SB1 systems (Norm-SB1). Lower-diagonal values mirror upper-diagonal ones. Dashes indicate self-comparisons of identical populations.}
    \label{Tab:KStest_LS_SB1}
    \resizebox{0.49\textwidth}{!}{ 
    \begin{tabular}{lcccccc} \hline \hline
                   &  N  & Run-LS & Run-SB1   & Norm-LS    & Norm-SB1\\ \hline
        Run-LS   & 53  & -         & 9\%         & 6\%         & 7\% \\
        Run-SB1  & 12  & 9\%       & -           & 21\%        & 7\% \\
        Norm-LS  & 64  & 6\%       & 21\%        & -           & 70\% \\
        Norm-SB1 & 39  & 7\%       & 7\%         & 70\%        & - \\ 
        \hline
    \end{tabular}
    }    
\end{table}

\end{appendix}

\end{document}